\documentclass[letter,11pt]{article}
\usepackage[utf8]{inputenc}
\usepackage[margin=1in]{geometry}
\usepackage{multirow}
\usepackage{nicefrac}
\usepackage{adjustbox}
\usepackage{mathtools} 
\usepackage{amssymb}
\usepackage{fdsymbol}
\usepackage{bm}
\usepackage{soul}
\usepackage{graphicx}
\usepackage{xcolor}
\usepackage{caption}
\usepackage{subcaption}
\usepackage{lineno}
\usepackage[english]{babel}
\usepackage{authblk}
\usepackage{booktabs}
\usepackage[square,numbers,sort]{natbib}
\usepackage{hyperref}
\graphicspath{{figs/}}

\newcommand{\on}{{\mathrm{on}}}
\newcommand{\JM}{{\mathrm{JM}}}
\newcommand{\off}{{\mathrm{off}}}
\newcommand{\rms}{{\mathrm{rms}}}
\newcommand{\Rey}{{\mathrm{Re}}}

\begin{document}
\title{Homogeneous turbulence in a random-jet-stirred tank}
\author[1]{Joo Young Bang}
\author[1]{Nimish Pujara}
\affil[1]{Department of Civil and Environmental Engineering, University of Wisconsin--Madison, Madison WI 53706, USA}
\date{}
\maketitle

\begin{abstract} 
We report an investigation into random-jet-stirred homogeneous turbulence generated in a vertical octagonal prism shaped tank where there are jet arrays on four of the eight vertical faces. We show that the turbulence is homogeneous at all scales in the central region of the tank that spans multiple integral scales in all directions. The jet forcing from four sides in the horizontal direction guarantees isotropy in horizontal planes but leads to more energy in the horizontal fluctuations compared with the vertical fluctuations. This anisotropy between the horizontal and vertical fluctuations decreases at smaller scales, so that the inertial and dissipation range statistics show isotropic behaviour. Using four jet arrays allows us to achieve higher turbulence intensity and Reynolds number with a shorter jet merging distance compared to two facing arrays. We show that we are able to vary the turbulence scales and Reynolds number by adding attachments to the exits of each jet and changing the parameters of the algorithm that drives random-jet stirring. By linking how the flow from each jet influences the random-jet-stirred turbulence, we provide recommendations on how to select the parameters of the jet driving algorithm.
\end{abstract}

%
\section{Introduction}\label{sec:introduction}
According to classical theories of turbulence, the statistical properties of turbulent motions whose size is small compared to scale of energy injection are universal, and locally homogeneous and isotropic \citep{Frisch_1996, pope2000turbulent, Davidson2015}. This motivates much research into fundamental and applied aspects of turbulent flows using `idealised turbulence' whose statistics are homogeneous and isotropic. For laboratory experiments, this has meant that the generation of idealised (homogeneous and isotropic) turbulent flow has been widely attempted in variety of stirred tank configurations. Different configurations include flow driven by oscillating grids \citep[\textit{e.g.},][]{DeSilva1994, Brunk1996, Blum2010, PoulainZarcos2022}, impellers or rotating disks \citep[\textit{e.g.},][]{Douady1991, Birouk1996, Voth1998, moisy1999kolmogorov, Worth2011, Lawson2015, dou2016piv, Bounoua2018, pujara2021measurements, lawson2022unsteady}, loudspeakers \citep[\textit{e.g.},][]{hwang2004creating, chang2012experimental, hoffman2021isotropic}, jets \citep[\textit{e.g.},][]{variano2008random, goepfert2010characterization, carter2016generating, johnson2018turbulent, tan2023scalings}, amongst other methods \citep[\textit{e.g.,}][]{Rensen2005}.

The different configurations to stir fluid into a turbulent state can be divided into three useful categories: (1) whether the stirring is continuous and steady (\textit{e.g.}, oscillating grids, steady rotations of disks and impellers, continuous actuation of jets) or unsteady and randomized (\textit{e.g.}, randomly actuated jets and impellers); (2) whether the stirring is provided by a single unit (\textit{e.g.}, a pair of counter-rotating disks, a set of oscillating grids) or multiple units (\textit{e.g.}, an array of jets or impellers); and (3) whether the stirring is asymmetric (\textit{e.g.}, an oscillating grid or jet array on one side of the tank) or symmetric (\textit{e.g.}, jets, impellers, or loudspeakers acting from multiple sides of the tank). It is generally found that continuous forcing is able to achieve higher Reynolds number turbulence compared to randomized forcing, but at the cost of stronger mean flows, higher mean shear, and a smaller volume of homogeneous isotropic turbulence \citep{voth2002measurement, hwang2004creating, Variano2004, Blum2010, Roy2012, pujara2021measurements}. Comparing single- and multi-unit forcing, multi-unit forcing produces a flow that is more complex to analyse and predict, but potentially allows more control over the scales of motion \citep{variano2008random, perez2016effect, carter2016generating, Bounoua2018}. Finally, while asymmetric forcing is unavoidable for certain setups \citep{variano2013turbulent, johnson2020sediment}, symmetric forcing produces better homogeneity and isotropy over a larger region with a smaller mean flow \citep{zimmermann2010lagrangian, goepfert2010characterization, bellani2014homogeneity, dou2016piv, hoffman2021isotropic}.

Here, we use flow data from a new facility where a column of water is stirred with four randomly actuated jet arrays arranged symmetrically around a vertical octagonal prism (see figure \ref{fig:schematic_diagram}a below) to gain a better understanding of how different aspects of multi-unit, unsteady and randomized stirring control the characteristics of turbulence produced. More specifically, we investigate how the turbulence statistics are influenced by the algorithm that controls the randomized stirring, the geometry of the jet arrangement, the size of the tank, and the properties of each jet. While such results have been previously reported from tanks with a single jet array \citep{variano2008random, perez2016effect, johnson2018turbulent} and two facing jet arrays \citep{bellani2014homogeneity, carter2016generating}, we address some outstanding questions and extend this knowledge to a new tank geometry and study the influence of individual jets in more detail. Based on a careful analysis of our data, we also provide recommendations on how to design and operate such facilities as well as how to analyse turbulence data from them. 

We begin in Sec.~\ref{sec:background} with a review of the existing knowledge from previous facilities to highlight the outstanding questions. Sec.~\ref{sec:experiments} describes the details of the apparatus and measurement, with the flow analysis detailed in Sec.~\ref{sec:flowanalysis}. A summary of the main results and implications for future research are given in Sec.~\ref{sec:discussion}.
%
\section{Background}\label{sec:background}
\paragraph{Algorithm for randomized stirring} 
The ``sunbathing'' algorithm introduced by \citet{variano2008random} randomizes the jet firing pattern in time and space to maximize generation of turbulent kinetic energy and minimize mean flow. An independent investigation \citep{perez2016effect} confirmed the superiority of this algorithm over others, though it is notable that slightly different algorithms have been found to be most effective in active grids \citep{Mydlarski1998} and a tank stirred with vertical rotating paddles where both directions of paddle rotation are equal in terms of driving flow towards the tank centre \citep{pujara2021measurements}. In the sunbathing algorithm, individual jets are turned on and off repeatedly, and the duration of each on/off period is chosen from Gaussian distributions described by $t_\on \sim N(\mu_\on,\sigma_\on^2)$ and $t_\off \sim N(\mu_\off,\sigma_\off^2)$, where $\mu_\on$ and $\mu_\off$ are the mean on/off times, and $\sigma_\on$ and $\sigma_\off$ are the standard deviations of the on/off times. Since flow statistics are insensitive to the standard deviations of the on/off times, the recommended value for the ratio of the standard deviations to the means of the on/off distributions is $\sigma/\mu = 1/3$. With this value chosen, the only remaining free parameters are the mean on-time $\mu_\on$ and the source fraction $\phi = \mu_\on/(\mu_\on+\mu_\off)$, which represents the mean of the fraction of the pumps firing at any given time. 

Investigations of the effects of the source fraction $\phi$ have found that its optimal value, where the root-mean-square velocities are maximised, is in the 5-25\% range for a single array \citep{variano2008random} and two facing arrays \citep{lawson2022unsteady}. The low values of the optimal $\phi$ relate to the fact that turbulence production is maximised by the interaction of the flow from individual forcing units with the background flow. Turning on all units robs the flow from each unit to both contribute to and interact with an unsteady background.

The source fraction $\phi$ is a dimensionless quantity, but the mean on-time $\mu_\on$ is generally reported as a dimensional quantity suggesting that the mechanism by which it influences the scales of turbulence is not fully understood. From systematic studies of varying $\mu_\on$ \citep{variano2008random, carter2016generating, perez2016effect, johnson2018turbulent}, it is known that increasing $\mu_\on$ increases the root-mean-square velocities and turbulence intensity until a certain value (different in each setup) at which point the effect saturates and further increases in $\mu_\on$ don't produce further increases in the flow intensity. For impeller arrays, \citet{lawson2022unsteady} introduced a dimensionless mean on-time based on the impeller rotation frequency and found that this value needs to be high ($O(10^3)$). For jets, there are two natural choices for making the mean on-time dimensionless:
\begin{equation}
    \frac{\mu_\on U_J}{D_J} \quad \text{or}\quad \frac{\mu_\on U_J}{L_A}, \label{eq:dimensionlessontime_options}
\end{equation}
where $U_J$ is the jet exit velocity, $D_J$ is the jet diameter, and $L_A$ is the array-to-array distance (or a representative scale for the tank size). The first choice ($\mu_\on U_J/D_J$) is analogous to the jet's `formation time' \citep{gharib1998universal} and postulates that the effects of mean on-time are related to the flow of an individual jet, while the second choice ($\mu_\on U_J/L_A$) relates the distance travelled by the fluid during a typical on-cycle to the size of the tank and postulates that it is whether the flow from each jet `reaches' the tank centre that matters. 

Another important question is whether there is a characteristic forcing timescale that determines the rate at which the largest scales of flow evolve. Based on the unsteady nature of the forcing in the sunbathing algorithm, several candidates have been proposed \citep{variano2008random, lawson2022unsteady}:
\begin{equation}
   \tau_{F1} = \mu_\on+\mu_\off = \frac{\mu_\on}{\phi}; \quad\quad \tau_{F2} = \mu_\on ; \quad\quad \tau_{F3} = \phi \mu_\on, \label{eq:forcingtimescales}
\end{equation}
where we have used the relation $\phi = \mu_\on/(\mu_\on+\mu_\off)$ in the definition of $\tau_{F1}$. These candidate forcing timescales all increase with $\mu_\on$, but differ on the role of $\phi$. 

\paragraph{Jet characteristics}
The flow characteristics of each forcing element are expected to affect the characteristics of turbulence in mult-unit forcing. For jets, the canonical non-swirling turbulent round jet is fully characterised by its Reynolds number $\Rey_J = U_J D_J/\nu$, where $\nu$ is the kinematic viscosity of the fluid, and it is thought that once $\Rey_J$ is high enough its value is less important than how the flow from different jets interact. However, since the flow from individual units is not typically characterised in detail, it is not clear what the characteristics are of each forcing element and whether manipulating them allows another way to control the turbulence produced.

\paragraph{Jet spacing and merging} 
Previous studies have hypothesized that as different jets fire and mix with the background flow, their `signature' is lost once they are fully merged. For continuously operating jets, scaling arguments from canonical non-swirling turbulent round jets can be used to predict the distance over which adjacent jets merge: The jets will merge when their half-widths intersect:
\begin{equation}
    2 S L_\JM = J \label{eq:jetmerging6J}
\end{equation}
Here, $L_\JM$ is the jet merging distance, $S$ is the jet half-width spreading rate and $J$ is the inter-jet spacing. Using $S \approx 0.09$ for a canonical round jet \citep{pope2000turbulent} gives $L_\JM \approx  5.5J$ \citep[for experimental evidence for this value, see][]{tan2023scalings}. In jet arrays with unsteady and randomized forcing, two main factors can be expected to cause deviation of $L_\JM$ from this value. First, the background (turbulent) flow affects the spreading rate $S$; \citet{khorsandi2013effect} and \citet{Guo_2005} find that jets in a turbulent environment are wider, their width grows faster compared to a quiescent background, and their axial velocities are arrested when the local background turbulence velocity is of the same order as the axial velocity. Second, since not all jets are always on, the effective inter-jet spacing is a function of the source fraction; this effective inter-jet spacing will be larger than $J$ though it is still expected to scale with $J$. While the first fact decreases the jet-merging distance, the second fact increases it.  

Evidence from a single randomly actuated jet array \citep[][]{variano2008random, perez2016effect, johnson2018turbulent} suggests that these two effects may roughly cancel each other out. It was found that the signature of individual jets could not be observed on the free surface above a depth of $6J$ for an upwards-pointing jet array placed at the bottom of a water tank. Similarly, profiles of root-mean-square velocities become uniform in directions parallel and perpendicular to the jet array at a distance of approximately $6J$ (before starting to decay as distance from the jet array increases further). For two jet arrays on opposite sides of the tank that face each other \citep[][]{bellani2014homogeneity, carter2016generating}, the data again support that $L_\JM \approx 6J$ is reasonable; the flow statistics become homogeneous at this distance from each jet array. However, it remains to clearly define what is meant by jet merging in randomly actuated jet arrays and it is unclear how sensitive the jet merging distance is to the source fraction and tank geometry.

\paragraph{Homogeneity, isotropy, and scales of turbulence}
Homogeneity and isotropy are easier to achieve for symmetric forcing \citep[\textit{e.g.}, two facing jet arrays or loudspeakers in all corners of a cube,][]{bellani2014homogeneity, hwang2004creating}. While forcing from many different directions appears to be a common approach when using continuous forcing \citep{hwang2004creating, hoffman2021isotropic, Bounoua2018}, planar symmetry (two facing arrays) is the most common configuration for randomized forcing \citep{bellani2014homogeneity, carter2016generating, lawson2022unsteady}. Two facing arrays with randomized forcing commonly produce a relatively large region of homogeneous turbulence, but only \citet{bellani2014homogeneity} seem to have achieved large-scale isotropy (measured by the ratio of root-mean-square velocities in the longitudinal and transverse directions with respect to the forcing). In \citet{carter2016generating}, the longitudinal (jet parallel) root-mean-square velocity is always higher than the transverse root-mean-square velocity, with 1.35 being the lower bound value for this ratio in their setup. Similarly, \citet{esteban2019laboratory} find a root-mean-square isotropy ratio of 1.2. \citet{bellani2014homogeneity} claim they achieve isotropy by optimizing the inter-array distance ($L_A$ in our notation) and it has been previously noted that the centrifugal nature of the pumps may play a role, but we hypothesize that the jet arrangement in their setup is also important. The pumps driving the jets are connected to 90 degree elbows that are insufficiently long to ensure the jet axis is perpendicular to the array, but because each group of four pumps is rotated relative to each other in the array plane, the net effect is flow isotropy in the tank centre. The other notable results from previous studies are that doubling the inter-jet spacing $J$ has little effect on the turbulence statistics and placing a mesh grid in front of each array reduces the turbulence intensity and integral scale in the tank centre \citep{carter2016generating}.

Interestingly, \citet{lawson2022unsteady} are able to produce homogeneous turbulence in a volume that is a significant fraction of the tank volume, and though their root-mean-square velocity isotropy ratio is of a similar magnitude to other facilities, it is the longitudinal velocity component that is weaker. Both of these effects appear to be related to the fact that, with impellers driven randomly in both directions, the forcing algorithm drives momentum towards the tank centre as well as back towards tank walls. This significantly reduces the merging distance and pumps more energy into the transverse component of velocity.

Apart from homogeneity and isotropy, it would be desirable to be able to predict the scales of turbulence based on a given tank design: the root-mean-square velocity (\textit{i.e.}, turbulent kinetic energy), the integral scale of turbulence, and the turbulent kinetic energy dissipation rate. As discussed above, previous results show that these quantities increase with mean on-time $\mu_\on$ (up to a saturation point) and vary non-monotonically with the source fraction $\phi$ (with a peak in the range 5-25\%), but it is not clear what sets the scales of turbulence in the first place and why there is a saturation point for them with increasing mean on-time. 
%
\section{Experiments} \label{sec:experiments}
%
\subsection{Turbulence tank}\label{sec:turbulencetank}
\begin{figure}
    \centering\
    \includegraphics[width=0.9\textwidth]{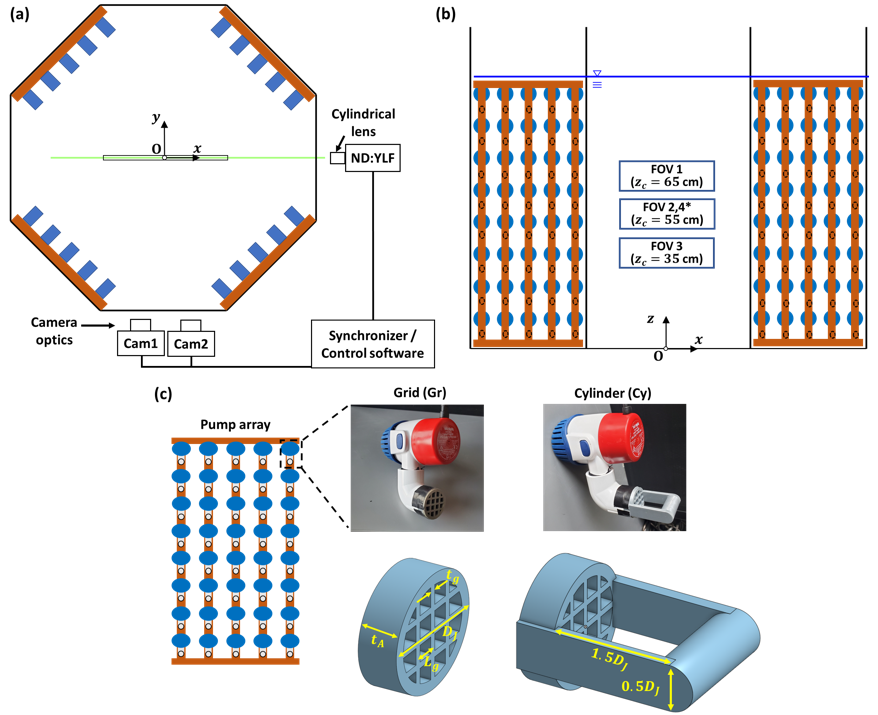}
    \caption{Schematic diagram of experimental apparatus: (a) top view, (b) side view, and (c) pump arrays with the grid and cylinder attachments. Particle image velocimetry (PIV) data is taken at four fields of view (FOVs). The positions of FOVs1--3 are $x=$ $\pm$ $16$ cm and $z=z_c$ $\pm$ $5$ cm at $y=0$. FOV4 is located at the same $x$ and $z$ position as FOV2, but in the $y = 7$ cm plane. The dimensions of the grid attachment are $2.66$ cm nozzle exit diameter ($D_J$), $1.8$ mm grid thickness ($t_g$), $\tfrac{1}{4} D_J$ grid spacing ($L_g$), and $1$ cm thickness ($t_A$). The solidity (blockage) of the grid attachment is $41.8$ \%. The cylinder attachment has the same design as the grid attachment but has a horizontal cylinder of diameter $\tfrac{1}{2}D_J$ placed a distance $\tfrac{3}{2} D_J$ in front of the nozzle exit.}
    \label{fig:schematic_diagram}
\end{figure}
The tank shown is an octagonal cylinder shape with a side of $61$ cm and a height of $122$ cm, constructed with acrylic plates supported by an aluminum frame (figure \ref{fig:schematic_diagram}). We place our coordinate system origin at the center of the tank bottom with the coordinates $x$ and $y$ comprising the horizontal plane and $z$ pointing in the vertical direction against gravity. During operation, the tank is filled with water (via a 20-micron filter) up to a depth of $1$~m  and kept in a temperature-controlled room such that the water temperature is 25.4 C with kinematic viscosity $\nu = 8.85 \times 10^{-7}$ m$^2$/s. Turbulence is generated by fluid forcing from four arrays of pumps on the vertical walls arranged so that there are two sets of orthogonal facing arrays. The distance between facing arrays is $L_{A} = 118$ cm, which also accounts for the attachments on each pump described below. Each array has 8 rows of 5 pumps (Rule 360 GPH bilge pump), and the spacing between adjacent pumps is $J = 12$ cm. The bottom row and two side columns are $J/2 = 6$ cm away from the wall ends, but the top row is 10 cm away from the water surface to ensure the base of the pumps is fully submerged. This configuration with mirror symmetry with the walls is to minimize secondary flow in analogy with oscillating grids \citep{fernando1993note}.

Each pump creates a synthetic jet by drawing water radially from its base and injecting it from the outlet. We connected 90-degree elbows so that the outlet directs the flow toward the tank center with a jet exit diameter of $D_J = 2.66$ cm. In preliminary experiments, we found that the elbow does not fully bend the flow and the jet velocity was not normal to the outlet face. It is also well known that bends generate secondary swirling flows. Thus, to straighten and condition the flow, we attach a 3D-printed grid to the outlet of each pump (figure \ref{fig:schematic_diagram}c).

When operated together, the pumps are controlled by solid-state relays (SSR-Rack48, Measurement Computing Corporation) which can open and close the circuit to 12 V power supplies. The circuits also have power distribution boards with 5 A fuses to protect the circuit and pumps. Two 96-channel digital input/output modules (PCIe-6509, National Instruments) are connected to the racks to trigger the relay modules. With this configuration, we can control individual pumps using custom MATLAB (Mathworks) code. To drive the pumps, we adopt the now-classical ``sunbathing'' algorithm of \citet{variano2008random} as described in Sec.~\ref{sec:background}. 
%
\subsection{Flow data collection}\label{sec:flowdata}
\paragraph{Random-jet-stirred turbulence}
We conduct velocity measurements under a range of different conditions. We test two source fractions: Low ($\phi = 6.25 \%$) and High ($\phi = 12.5 \%$). While $\phi=12.5\%$ was found to be the optimal value for a single jet array \citep{variano2008random}, it is possible that the optimal value shifts to lower values as more jet arrays are added and the total number of jets increases. We also test three different mean on-times: $\mu_\on = [1.5, 3, 6]$ s to understand the growth and saturation of turbulence intensity as a function of the mean on-time. Finally, we test the effect of attachments at the pump outlets to understand how manipulations of individual forcing elements affect the scales of turbulence. The pumps with grid attachments are considered the baseline case and the pumps with a grid and a horizontal cylinder attachment are the new test case (figure \ref{fig:schematic_diagram}c). We hypothesized that including a horizontal cylinder would weaken the jet and increase vertical stirring via flow separation around the cylinder, both of which should contribute to improved homogeneity and isotropy of the turbulence. In all, there are 12 different experimental conditions, as summarized in table \ref{table:experimental_conditions}. 
\begin{table}
\caption{Experimental conditions.}
\centering
\begin{adjustbox}{width=0.8\textwidth}
\small
\centering
\begin{tabular}{|c|c|c|c|c|c|c|} 
\hline
Case & Attachment & $\phi$ (\%)& $\mu_\on$ (s) & $\mu_\off$ (s)& $\sigma_\on$ (s)& $\sigma_\off$ (s) \\
\hline 
GrLo15 & \multirow{6}{*}{Grid} & 6.25 & 1.5 & 22.5 & 0.5 & 7.5 \\
GrLo30 &  & 6.25 & 3.0 & 45.0 & 1.0 & 15.0 \\
GrLo60 &  & 6.25 & 6.0 & 90.0 & 2.0 & 30.0 \\
GrHi15 &  & 12.5 & 1.5 & 10.5 & 0.5 & 3.5 \\
GrHi30 &  & 12.5 & 3.0 & 21.0 & 1.0 & 7.0 \\
GrHi60 & & 12.5 & 6.0 & 42.0 & 2.0 & 14.0 \\
\hline
CyLo15 & \multirow{6}{*}{Cylinder} & 6.25 & 1.5 & 22.5 & 0.5 & 7.5 \\
CyLo30 &  & 6.25 & 3.0 & 45.0 & 1.0 & 15.0 \\
CyLo60 &  & 6.25 & 6.0 & 90.0 & 2.0 & 30.0 \\
CyHi15 &  & 12.5 & 1.5 & 10.5 & 0.5 & 3.5 \\
CyHi30 &  & 12.5 & 3.0 & 21.0 & 1.0 & 7.0 \\
CyHi60 &  & 12.5 & 6.0 & 42.0 & 2.0 & 14.0 \\
\hline
\end{tabular}
\end{adjustbox}
\label{table:experimental_conditions}
\end{table}

Velocity fields are measured with 2D-planar particle image velocimetry (PIV) (figure \ref{fig:schematic_diagram}): A 527 nm Nd-YLF laser (Photonics Industries) with a cylindrical lens located at $+x$ side wall creates a light sheet with an average thickness of 1.5 mm in the $x$-$z$ plane. Images are taken with cameras (Phantom VEO430; 2560 px $\times$ 1600 px with 10 $\mu$m px size) installed at $-y$ side wall. Each camera is mounted with a 100 mm lens (Tokina) and fitted with a 527 nm bandpass filter. The flow is seeded with tracer particles (10 $\mu$m median diameter hollow microspheres with a specific gravity of 1.10 $\pm$ 0.05; Potter Industries 110P8). 

To examine the homogeneity of the turbulent flow in the vertical extent, we took velocity data at three different field of views (FOVs) (figure  \ref{fig:schematic_diagram}b). Using two cameras side-by-side, we captured velocity data covering $x=$ $\pm$ 16 cm and $z=z_c$ $\pm$ 5 cm, where $z_{c}$ is the vertical center of each FOV, at $y=0$. To check that the flow was homogeneous away from the centre plane, we also took data at FOV4, which was located at the same $x$-$z$, but at $y=7$ cm. The magnification factor in these images is approximately 14 px/mm, and the time interval between image pairs is adjusted to give a maximum particle displacement of 5-7 pixels. At each FOV, we took 1460 image pairs with a 0.5 Hz sampling rate and for certain cases, we also took 1460 image pairs with a 20 Hz sampling rate in order to resolve higher frequencies.

Image processing to obtain velocity fields is conducted using DaVis v10 (LaVision). Since particle intensities can vary significantly across the FOV, we use a nonlinear filter to match the dynamic range of the particle intensities within each image \citep{adrian2011particle}, which significantly improves the signal-to-noise ratio. Velocity vectors are calculated using an iterative cross-correlation method with four passes for each image pair. The first 64 $\times$ 64 px pass is followed by three passes at 32 $\times$ 32 px with 75 \% overlap. The spatial resolution is about 3 to 5 times the Kolmogorov length scale, which in turn, is sufficient to capture more than 95\% of the turbulent kinetic energy \citep{saarenrinne2001experiences}. Sub-pixel accuracy in the pixel displacement in each sub-window is achieved using a Gaussian fitting function to the correlation maximum and neighboring pixels. Each velocity vector is quality-checked by comparing the relative heights of the first and second peaks in the correlation field. We achieve more than 95\% valid vectors and low-quality vectors are removed (without interpolation) from the analysis.

\paragraph{Single jet flow}
To complement velocity measurements of turbulence driven by randomly actuated jets, we also conduct experiments where the flow due to a single pump is measured. A single pump is positioned at the $-x$ side wall so that the light sheet bisects the pump's outlet and allows for taking data of the pump jet's velocity profile. We take data of the pump with the grid attachment and with the cylinder attachment, in which case the cylinder axis is aligned with the $y$ axis. We measure the jet's velocity profile in continuous mode and in pulsed mode with pulse durations of $1.5$, $3$, and $6$ s to mirror the mean on-times used to generate turbulence. Velocity data is taken in the jet's near field ($x = 5 D_J$) and in the tank centre ($x = \tfrac{1}{2}L_{A} \approx 22.2 D_J $). 
%
\section{Flow analysis} \label{sec:flowanalysis}
%
\subsection{Velocity from a single jet}\label{sec:singlejetdata}
\begin{figure}
    \centering
    \includegraphics[width=0.6\textwidth]{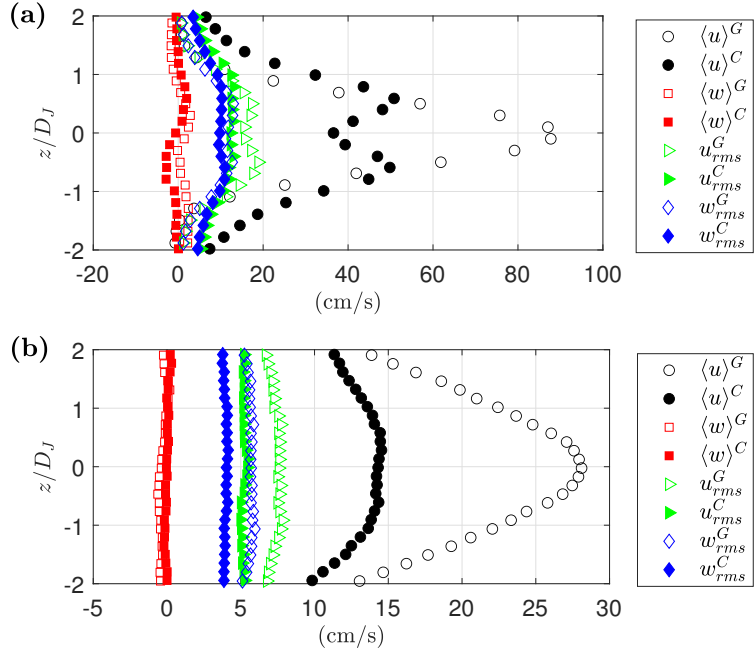}
    \caption{Velocity profiles of continuous jet flow with the grid attachment and cylinder attachment in the near field $x=5D_J$ (a) and at the tank centre $x=\tfrac{1}{2}L_{A}$ (b). The superscripts refer to the grid (G) and cylinder (C) attachments, respectively.} 
    \label{fig:5D_TC_Gr_Cy_continuous_jet}
\end{figure}
The velocity data from the single pump operated continuously is decomposed into a temporal mean and fluctuating component, with root-mean-square velocities calculated from the fluctuating component. The profiles of the mean and fluctuating components are shown in figure \ref{fig:5D_TC_Gr_Cy_continuous_jet}. We find that the grid attachment successfully conditions the flow so that the resulting jet resembles a classical non-swirling round jet with flow normal to the outlet. On the other hand, the cylinder attachment creates a wake at the jet centreline in the near-field (figure \ref{fig:5D_TC_Gr_Cy_continuous_jet}a), but this velocity deficit is smoothed out into a weaker and broader jet relative to the grid attachment by the time the jet reaches the tank centre (figure \ref{fig:5D_TC_Gr_Cy_continuous_jet}b). 

From the near-field data (figure \ref{fig:5D_TC_Gr_Cy_continuous_jet}a), we also make estimates of the jet-exit velocity $U_J$ by taking the peak axial velocity. This gives
\begin{subequations}
\begin{align}
U_J &=0.90 \text{ m/s for the grid attachment}, \\
U_J &=0.50 \text{ m/s for the cylindrical attachment}.
\end{align}
\end{subequations}
For the grid attachment, this value is within 5\% of the value obtained from fitting the far-field data to scaling relations for canonical round jets \citep[Eq. (5.6) in][]{pope2000turbulent}, whereas for the cylindrical attachment, the far-field data do not show the scaling in canonical round jets. Thus, using the peak axial velocity in the near-field gives a consistent estimate of $U_J$ for both attachments. 

\begin{figure}
    \centering
    \includegraphics[width=0.5\textwidth]{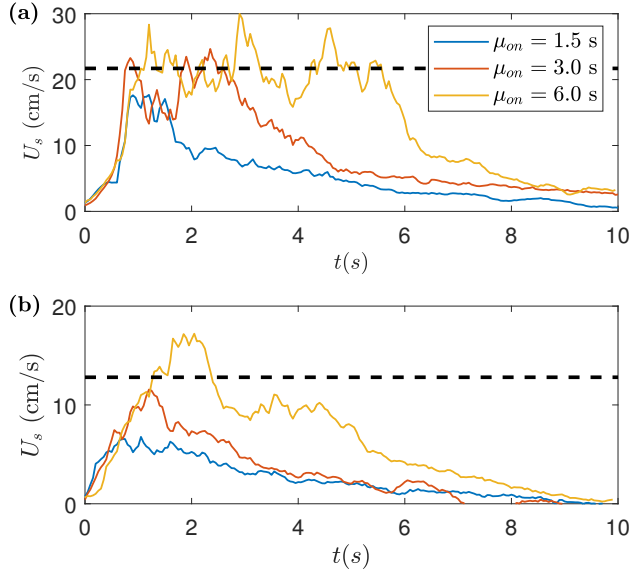}
    \caption{Temporal evolution of the spatial average axial velocity ($U_{s}$) with the pulse firing with the grid (a) and cylinder (b) attachments. The spatial averaged is linearly conducted over $z/D_J = \pm 2$ at $x=L_{A}/2$. Dashed lines represent $U_{s}$ calculated from continuous jet flow experiments.} 
    \label{fig:Us_pulsed_TC}
\end{figure}
We next consider the velocity data from the single jet operated in pulsed mode, where the pulse lengths are 1.5 s, 3 s, or 6 s (to mirror the mean on-times used in the sunbathing algorithm to produce random-jet-stirred turbulence). To compare the time evolution of the axial velocity field across different on-times, we calculate the velocity $U_s$, which is the averaged velocity over $z/D_J = \pm 2$ at the tank centre. The results in figure \ref{fig:Us_pulsed_TC} show that the velocity magnitude in pulsed mode is smaller than in continuous mode, until the pulse length is long enough after which it is essentially the same as that in a continuous jet. 

The reduced axial velocity for short pulse lengths compared to continuous jets is related to several factors: (1) The ramp-up time of the pumps, which is a greater fraction of the pulse duration for short pulse lengths; (2) The vortex ring that is created during the jet start-up when the flow separates from the nozzle in a spiral roll-up \citep{gharib1998universal}. We detect this vortex ring for both attachments and all pulse durations; and (3) The diffusion of axial momentum as the pulsed jet flow travels to the tank centre. The ramp-up time is short compared to our pulse lengths, so it is not expected to play a significant role. For the starting vortex, it is known that such a vortex ring absorbs the momentum of the discharged fluid and grows in size even after it detaches from the nozzle by absorbing momentum from the trailing jet \citep{Schram_2001_jet, gao2010model}. However, for a jet formation time $T_\on U_J /D_J \gg 4$ \citep{gharib1998universal}, there is a clear trailing jet. This criteria is satisfied for all pulse lengths tested here. Thus, the most important effect must be the diffusion of axial momentum, which is a function of how far the fluid must travel ($\tfrac{1}{2}L_A$ in this case). From the data in figure \ref{fig:Us_pulsed_TC}, we find that the measured axial velocity for a pulsed jet is close to the continuous jet value when $T_\on U_J /(\tfrac{1}{2}L_A) \gtrsim 5$. 
%
\subsection{Random-jet-stirred turbulence}\label{sec:randomjetstirred}
\begin{figure}
    \centering
    \includegraphics[width=\textwidth]{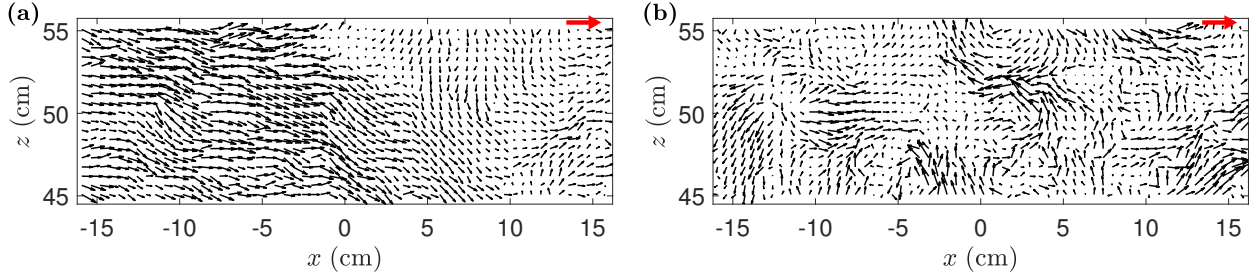} 
    \caption{Instantaneous velocity snapshots at FOV2 of the GrLo30 case corresponding to the maximum (a) and minimum (b) instantaneous kinetic energy. For visual clearance, only every 10$^\text{th}$ vector is shown. The red arrows show a magnitude of $40$ cm and $10$ cm/s, respectively, for reference.}
    \label{fig:snapshots_qmax_qmin}
\end{figure}

Figure \ref{fig:snapshots_qmax_qmin} shows instantaneous snapshots of the velocity field at FOV2 for the GrLo30 case. The flow fields, which  correspond to snapshots with the maximum and minimum instantaneous kinetic energy, exhibit rotation and shear at various scales that are the hallmarks of turbulent flow.
To conduct further analysis, we subject the velocity data to a Reynolds decomposition where the instantaneous velocity field $\bm{u} = (u, v, w) = (u_1,u_2,u_3)$, which is aligned with $(x, y, z)$ coordinates, is decomposed into an ensemble averaged velocity $\langle {\bm{u}} \rangle$ and fluctuating velocity ${\bm{u}}^\prime$
\begin{equation}
{\bm u} ({\bm x}, t) = \langle {\bm{u}} \rangle + {\bm u}^{\prime},
 \label{eq:velocity_decomposition}
\end{equation}
where the ensemble average $\langle\cdot\rangle$ is computed over all velocity snapshots. From the fluctuating velocity field, we compute root-mean-square (rms) of the fluctuations in a given component $u_\rms$ = $[{\langle u'^2 \rangle}]^{1/2}$ and the turbulent kinetic energy (TKE) $k = ({\langle u'^2 \rangle}+{\langle v'^2 \rangle}+{\langle w'^2 \rangle})/2 \approx (2{\langle u'^2 \rangle}+{\langle w'^2 \rangle})/2$. The approximation in computing $k$ comes from assuming horizontal isotropy. Since we force the flow from four orthogonal horizontal directions, we expect that flow statistics are invariant to rotations about the $z$ axis (\textit{i.e.}, $u_\rms = v_\rms$). In the subsequent analysis, we also use a spatial average $\overline{\cdot}$, which is computed over space in each FOV in the central region of the tank, $\lvert x \rvert < 10$ cm. Within this region, the statistics of turbulence are homogeneous, as shown below.
\begin{figure}
    \centering
    \includegraphics[width=\textwidth]{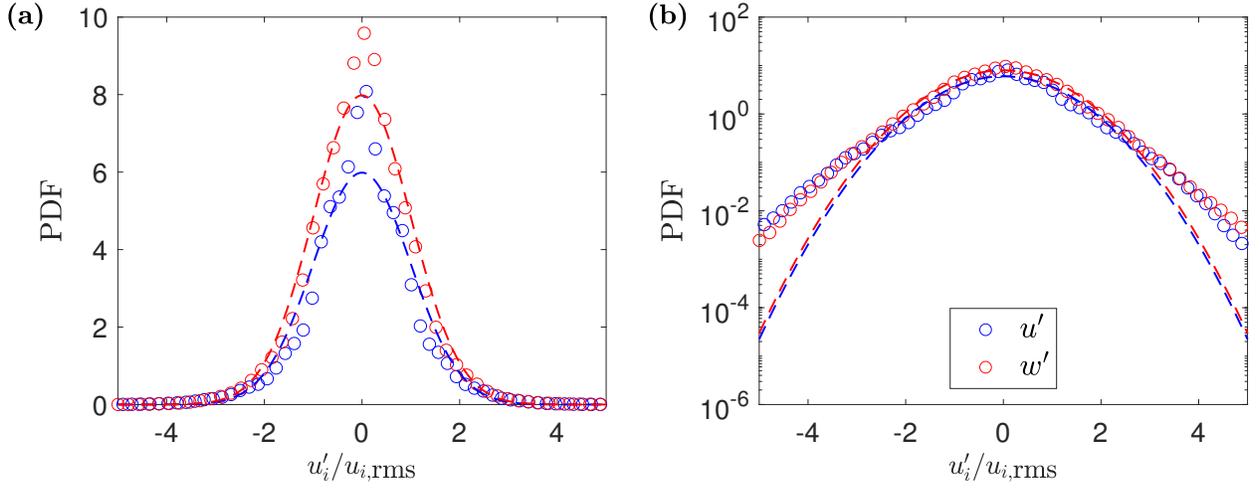}
    \caption{The normalized PDFs of fluctuating velocity at FOV2 of GrLo30 case plotted in linear (a) and log (b) scales. The dashed lines indicate Gaussian distributions obtained using $u_{i,\rms}$.} 
    \label{fig:PDF_GrLo30_F2}
\end{figure}

The normalized probability density functions (PDF) of the fluctuating velocities in the homogeneous region at FOV2 for the GrLo30 case are displayed in figure \ref{fig:PDF_GrLo30_F2} with Gaussian distributions using their standard deviation. The moments of $\bm{u}^\prime$ in the homogeneous region at FOV2 are also summarized in table \ref{table:1pt_TurbulenceStatisticsFOV2}. The skewness is always negligible, but the kurtosis is always greater than 3 indicating that extreme values in the fluctuating velocities are more likely than in a Gaussian distribution, which can also be seen in the tails of the PDFs in figure \ref{fig:PDF_GrLo30_F2}b. The tails are presumably related to the jet stirring. 
\begin{table}
\caption{One-point turbulence statistics in the homogeneous region at FOV2}
\begin{adjustbox}{width=1\textwidth}
\small
\centering
\begin{tabular}{|c|c c c c|c c c c| c c c|}
\hline
Case \# & $\overline{\langle u \rangle}$ & $\overline{\langle w \rangle}$ & $\overline{u_\rms}$ & $\overline{w_\rms}$ & $S(u^\prime)$ & $S(w^\prime)$ & $K(u^\prime)$ & $K(w^\prime)$& $\overline{u_\rms /w_\rms}$ & $M_1$ & $M_2$\\
\hline
& \multicolumn{4}{c|}{cm/s} & \multicolumn{4}{c|}{} &\multicolumn{3}{c|}{}\\
\hline
GrLo15 & -0.31 & -0.21 & 5.78 & 4.52 & -0.03 & -0.07 & 4.38 & 4.01 & 1.28 & 0.052 & 0.004 \\
GrLo30 & -0.26 & -0.42 & 6.66 & 5.00 & -0.11 & -0.02 & 4.52 & 4.06 & 1.33 & 0.052 & 0.003 \\
GrLo60 & -0.13 & -0.44 & 7.11 & 5.03 & -0.05 & -0.04 & 4.50 & 4.26 & 1.41 & 0.038 & 0.002 \\
GrHi15 & -0.71 & 0.16 & 5.94 & 4.75  & -0.13 & -0.03 & 4.06 & 3.62 & 1.25 & 0.097 & 0.014 \\
GrHi30 & -0.38 & -0.07 & 7.47 & 5.91 & -0.08 & -0.03 & 3.72 & 3.47 & 1.26 & 0.046 & 0.003 \\
GrHi60 & -0.64 & -0.30 & 8.39 & 6.41 & 0.01 & -0.03 & 3.57 & 3.38 & 1.31 & 0.069 & 0.006 \\
\hline 
CyLo15 & -0.21 & -0.52 & 3.65 & 2.94 &-0.17 & -0.17 & 4.43 & 3.88 & 1.23 & 0.097 & 0.013\\
CyLo30 & 0.09 & -1.24 & 4.70 & 3.64 & 0.05 & -0.17 & 4.12 & 3.70 & 1.29 & 0.122 & 0.029\\
CyLo60 & -0.10 & -1.16 & 5.22 & 3.84 & -0.07 & -0.26 & 4.08 & 3.71 & 1.36 & 0.111 & 0.022 \\
CyHi15 & -0.56 & 0.51 & 3.61 & 3.02 & -0.12 & -0.11 & 4.23 & 3.61 & 1.18 & 0.162 & 0.031\\
CyHi30 & -0.06 & -0.79 & 4.76 & 3.87 & -0.07 & -0.08 & 3.81 & 3.34 & 1.23 & 0.093 & 0.013 \\
CyHi60 & -0.32 & -1.16 & 5.66 & 4.36 & -0.03 & -0.03 & 3.57 & 3.32 & 1.30 & 0.115 & 0.020 \\
\hline
\end{tabular}
\end{adjustbox}
\label{table:1pt_TurbulenceStatisticsFOV2}
\end{table}
%
\subsubsection{The homogeneous turbulence region and jet merging}\label{sec:homogeneity} 
\begin{figure}
    \centering
    \includegraphics[width=1\textwidth]{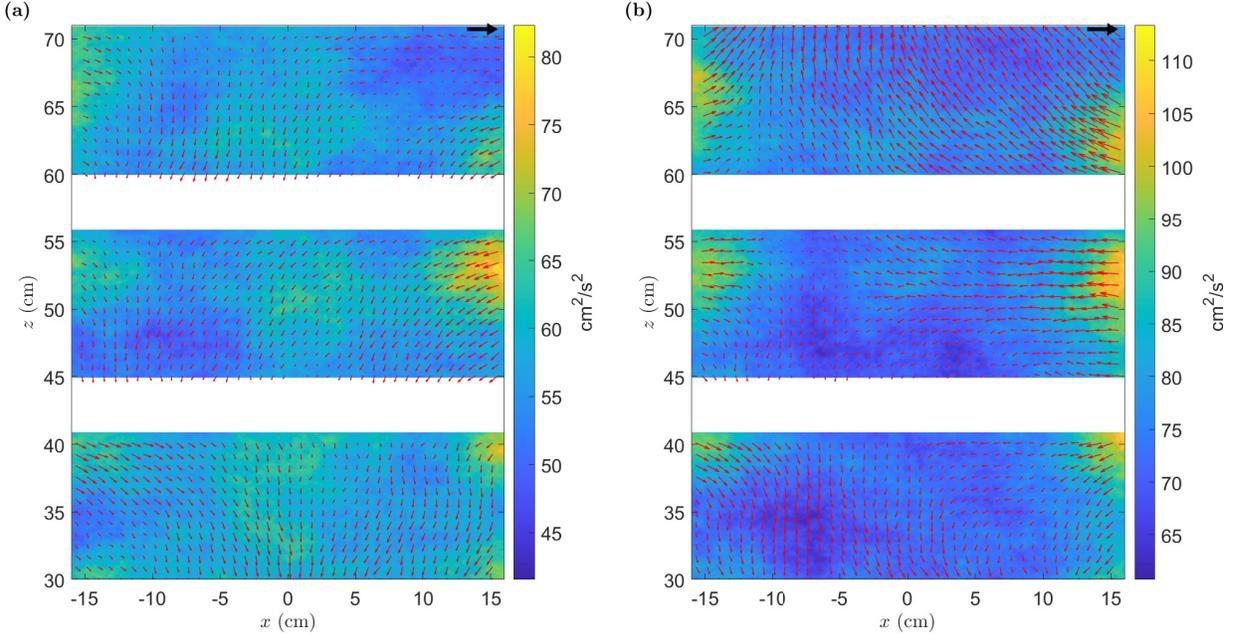}
    \caption{Ensemble averaged velocity and TKE fields for GrLo30 (a) and GrHi30 (b) over FOVs1--3 ($y=0$ plane). The large black arrow shows show a magnitude of $2$ cm/s.}
    \label{fig:mean_TKE_GrLoHi30}
\end{figure}
Figure \ref{fig:mean_TKE_GrLoHi30} presents the ensemble-averaged velocity and TKE fields for FOVs1--3 for the GrLo30 and GrHi30 cases. Examining the TKE fields, we note that it is relatively homogeneous in the $\lvert x \rvert <10$ cm region within each FOV. Specifically, the standard deviation of TKE in space is less than 5\% of its spatial mean $\overline{k}$, and moreover, $\overline{k}$ has very similar values across FOVs. Additionally, for $\lvert x \rvert > 10$, we see local patches of increased TKE at $z \approx 42, 54,$ and $66$ cm, which are the heights of the jets, but these `jet signatures' are absent for $\lvert x \rvert < 10$. This is true for both the low and high source fraction data shown in the figure.

Defining the jet merging distance $L_\JM$ as the distance from the jet array to the point where turbulent statistics are homogeneous, the data show that the value is not sensitive to the parameters of the sunbathing algorithm ($\phi$ and $\mu_\on$) lending support to the idea that $L_\JM \sim J$ (Eq. (\ref{eq:jetmerging6J})). However, we observe homogeneous statistics at a shorter distance ($L_{JM} \approx 4J$) than the previously accepted value of jet merging distance ($L_{JM} \approx 6J$). We attribute this shorter jet merging distance to the presence of the additional arrays. While jet flows from single or two facing arrays first mix with adjacent jets from the same array, the presence of jet arrays oriented perpendicular to each other in our setup allows mixing of jets across arrays allowing mixing at shorter distances.
%
\subsubsection{Large-scale isotropy and mean shear}\label{sec:isotropy} 
\begin{figure}
    \centering
    \includegraphics[width=0.65\textwidth]{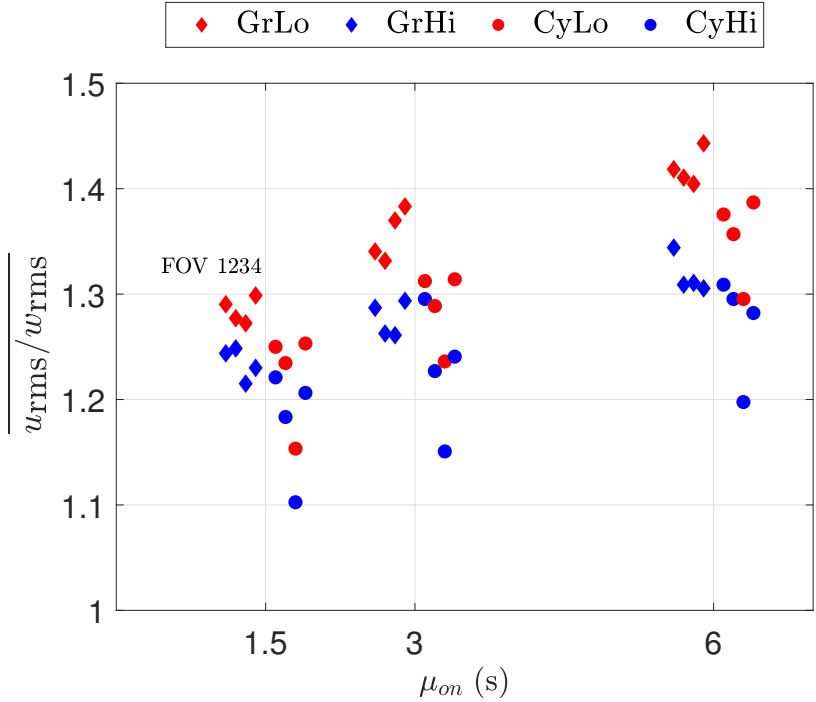}
    \caption{Spatially averaged RMS velocity ratio as a function of $\mu_\on$ for the grid attachment (diamonds) and the cylinder attachment (circles). Data for FOVs1--4 are displayed by offsetting them with respect to each other (left to right), as the text appears in both plots.}
    \label{fig:ISO_Gr_Cy}
\end{figure}

The geometry of the tank setup ensures that flow statistics are isotropic in the horizontal directions. To investigate large-scale isotropy of the flow in the vertical \textit{vs.} horizontal directions, figure \ref{fig:ISO_Gr_Cy} shows the spatially averaged ratio of the RMS velocities, which should be unity for large-scale isotropy. We observe values in the range 1.21--1.44 for the grid attachment and 1.1 -- 1.39 for the cylinder attachment. Higher source fraction $\phi$ and smaller mean on-time $\mu_\on$ give better performance in terms of achieving large-scale isotropy, consistent with previous results  \citep{variano2008random, perez2016effect, carter2016generating, johnson2018turbulent}. These trends can be understood by considering how an individual jet interacts with the existing background turbulence that it is fired into. While flow from each jet has the majority of its momentum in the jet-axial direction, the background turbulence that the jet travels into serves to break up the jet structure. For larger $\mu_\on$, a larger patch of fluid that retains the anisotropic signature of the jet flow is set into motion whereas for larger $\phi$, this patch must contend with a higher intensity background flow which breaks it up more effectively resulting in more isotropic statistics.

The cylindrical attachment results in only a slight improvement in the large-scale isotropy, despite the weakened and wider velocity profiles observed in the single jet data (figures \ref{fig:5D_TC_Gr_Cy_continuous_jet} and \ref{fig:Us_pulsed_TC}). The reason is that while there is clear evidence of flow separation and more vertical mixing in the jet's near field, the far field statistics such as $\langle u \rangle/w_\rms$ or $u_\rms/w_\rms$ are found to be similar to the grid attachment. It appears that since jets with manipulated exit conditions still relax to have properties analogous to canonical jets in the far field, the improvements in isotropy are small.

Another metric of large-scale isotropy is Reynolds stress tensor, where the off-diagonals are zero for isotropic turbulence.  In other words, $u^\prime$ and $w^\prime$ should be uncorrelated and no mean shear should exist. To evaluate the correlation between $u^\prime$ and $w^\prime$, the normalized Reynolds stress ($\mathrm{NRS}$) is calculated by
\begin{equation}
    \mathrm{NRS} = \overline{\left(\frac{|\langle u^\prime w^\prime \rangle|}{u_{\rms}w_{\rms}}\right)}
    \label{eq:NRS}
\end{equation}
The absolute value is applied prior to spatial averaging in order to prevent the Reynolds stress from being canceled out due to the different signs. The normalized Reynolds stress is less than 5 \% for the grid and 7 \% for the cylinder cases, suggesting that there is negligible correlation between velocity fluctuations and hence negligible TKE production in the tank centre.
%
\subsubsection{Mean flow structure and magnitude}\label{sec:meanflow}
Examples of the mean flow field ($\langle {u} \rangle, \langle {w} \rangle$) are shown in figure \ref{fig:mean_TKE_GrLoHi30}. While the mean flow is always small compared to the fluctuating flow field, we find that its structure depends on the parameters of the sunbathing algorithm. The $\langle {u} \rangle$ fields generally show inward-directed flows due to how the jets are directed, but the $\overline{\langle w \rangle}$ fields change sign depending on the parameters of the sunbathing algorithm. For example, in the grid attachment cases, $\overline{\langle w \rangle}$ is positive (upward directed) at FOV1, whereas $\overline{\langle w \rangle}$ is negative (downward directed) with the cylinder attachment at $\mu_\on = 3$ and $6$ s at FOVs2--3. From continuity considerations, the jet flows directed horizontally toward the tank centre must recirculate back towards the arrays. One would expect this recirculation to occur at the top and bottom of the tank where there are no arrays with a preference for the top of the tank where there is reduced friction due to the free surface. This would suggest upward directed mean flow in the tank centre, but at times we also observe downward directed mean flow and preferential return via the bottom of the tank. One possible explanation for this is the small amount of heat generated by the submerged pumps that could induce flow upwards adjacent to the pumps, which then sets a mean flow structure with downward vertical velocities. This mechanism is likely to be important in all tanks where the flow driving mechanism (e.g., pump, motors) is not thermally isolated from the working fluid.

To demonstrate that the mean flow is weak compared to the turbulence in the homogeneous region, we use the metrics
\begin{subequations}
\begin{align}
    M_1 =\overline{\left(\frac{2|\langle u \rangle|+|\langle w \rangle|}{2u_\rms+w_\rms}\right)} \label{eq:M1}\\
    M_2 =\overline{\left(\frac{2\langle u \rangle^2+\langle w \rangle^2}{2u^2_\rms+w^2_\rms}\right)}.\label{eq:M2}
\end{align}
\end{subequations}
$M_1$ compares the magnitude of the mean flow to the turbulence RMS velocity, and the absolute value of the mean velocity is used to avoid excessive minimization of $M_1$ by the spatial averaging. $M_2$ represents the ratio of the kinetic energy of the mean flow to the turbulence. Using the cutoff value of ${M_1} < 0.1$, we find four cases (GrLo15, 30, 60, and GrHi30) that satisfy this criterion for all FOVs. For these cases, $M_2<0.01$ (\textit{i.e.}, the TKE is more than 100 times larger than the kinetic energy of the mean flow). Surprisingly, all cases with the cylindrical attachment have ${M_1} > 0.1$. A closer examination reveals that mean flows for both attachments have relatively similar magnitudes, but the cases with the cylindrical attachment have smaller values of $\bm{u}_\rms$ leading to higher values of $M_1$ and $M_2$. 

Overall, the mean flow magnitude is not a function of the jet attachments and the parameters of the sunbathing algorithm, but its structure is. However, $M_1$ and $M_2$ are always less than 0.2 and 0.05, respectively, showing the mean flow is weak compared to the turbulence fluctuations. 
%
\subsubsection{Timescale of unsteady forcing} \label{sec:forcingtimescales}
As noted in Sec.~\ref{sec:background} and Eq. (\ref{eq:forcingtimescales}), there are different candidates for which timescale associated with the sunbathing algorithm most strongly affects the large-scale unsteady motions that decay into turbulence in the tank centre. We investigate this using the power spectra of the one-point velocity fluctuations ($u^\prime$), which are aligned with the jet axes of two of the four arrays. To compute the spectra, we first interpolate any missing velocity vectors using a spatial median filter with a window size of 3. The entire flow fields are then decomposed into 5 subsamples, each containing 292 flow fields. The final spectra are obtained by averaging the spectrum at each point over the homogeneous turbulence region, both within and across FOVs. This ensemble averaging allows us to reduce the statistical uncertainty. To cover different frequency ranges, we use velocity data acquired at 0.5 Hz and at 20 Hz.
\begin{figure}
    \centering
    \includegraphics[width=\textwidth]{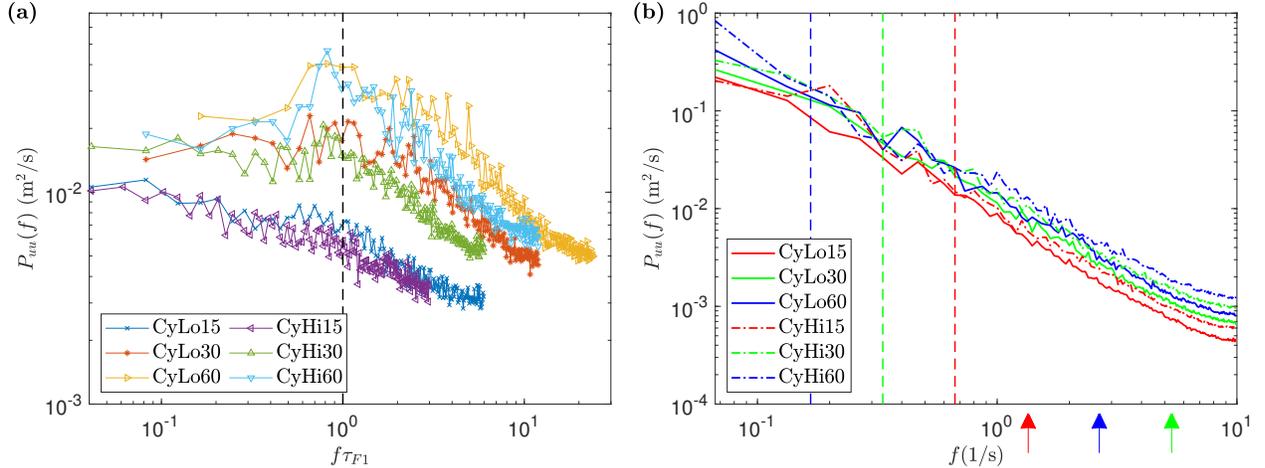}
    \caption{Power spectral density of $u^\prime$ for the cases with the cylinder attachment using 0.5 Hz velocity data (a) and 20 Hz velocity data (b). In panel (a), the frequency domain is normalized by $\tau_{F1}$ with the black dashed line showing $f\tau_{F1} = 1$. In panel (b), the spectra of the same $\mu_\on$ are displayed in the same colour and dashed coloured lines represent $f=(\tau_{F2})^{-1}$  and the arrows represent $f=(\tau_{F3})^{-1}$ for CyHi60 (red), CyHi30 and CyLo60 (blue), and CyHi15 and CyLo30 (green). The arrow related to the CyLo15 case ($f=10.67 s^{-1}$) is slightly beyond the frequency domain. Note, the noise levels in both panels is similar but the spectra in panel (a) appear noisier due to a smaller range of values.}
    \label{fig:PSD_Cy}
\end{figure}

Figure \ref{fig:PSD_Cy} shows the spectra for the cylinder attachment cases for velocity data taken at 0.5 Hz and 20 Hz in panels (a) and (b), respectively. No peaks are observed at frequencies associated to $\tau_{F2} =  \mu_\on$ and $\tau_{F3} = \phi \mu_\on$ (figure \ref{fig:PSD_Cy}b) suggesting that neither of these are the correct candidate for the forcing timescale. Additionally, no clear peaks are observed at frequencies associated to $\tau_{F1} = \mu_\on+\mu_\off$ for $\mu_\on = 1.5$ and $3$ s, but there is a convincing peak at this location for $\mu_\on = 6$ s. From this data, it appears that the correct timescale for unsteady forcing using the sunbathing algorithm is $(\mu_\on+\mu_\off)$ \citep[as suggested by][]{lawson2022unsteady} and not $\phi \mu_\on$ \citep[as suggested by][]{variano2008random}. However, the appearance of the spectral peak only at the highest value of $\mu_\on$ supports the argument of \citet{variano2008random} that there is a threshold $\mu_\on$ beyond which increases in the magnitude of the fluctuating velocity are signatures of the jet forcing and not turbulent motions. Based on the pulsed single jet data (Sec.~\ref{sec:singlejetdata}), we can relate this finding to the dimensionless mean on-time proposed in Eq.~\ref{eq:dimensionlessontime_options}. A $\mu_\on = 6s$ for the grid attachment translates to $\mu_\on U_J / (\tfrac{1}{2}L_A) \approx 10$. It appears that this is the threshold value for observing clear signatures of the forcing timescale in the flow at the tank centre. 
%
\subsubsection{Scales of motion: RMS velocities}\label{sec:scalesofmotion_velocity}
\begin{figure}
    \centering
    \includegraphics[width=0.7\textwidth]{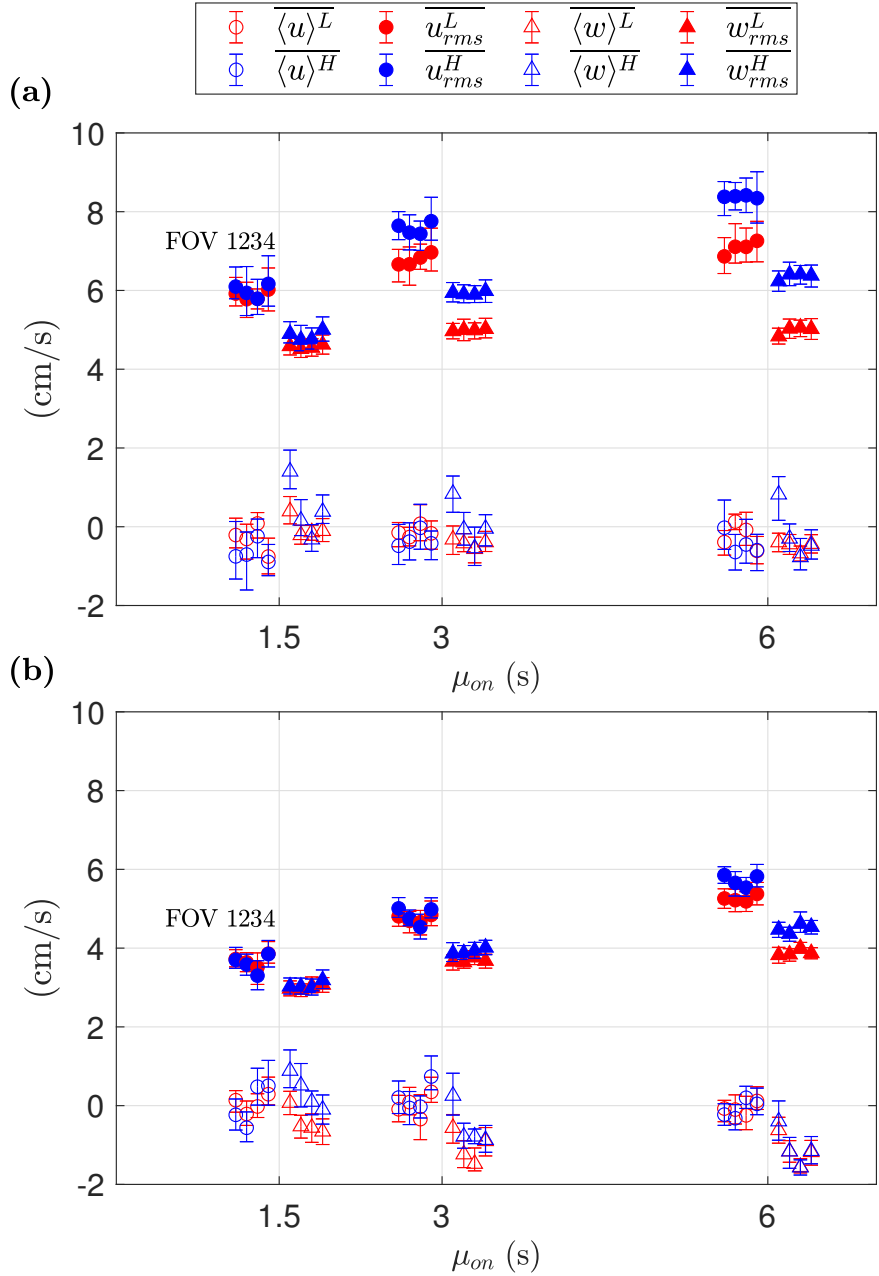}
    \caption{Summary of one-point statistics in the homogeneous region $\lvert x \rvert < 10$ cm as a function of $\mu_\on$ for the grid attachment (a) and the cylinder attachment (b). The symbol colour indicates low ($\phi = 6.25 \%$) and high ($\phi = 12.5 \%$) source fraction of pumps that are on at a given time. The error bars illustrate the spatial variability of each quantity, displaying 95 \% of the range. Data for FOVs1--4 are displayed by offsetting them with respect to each other (left to right), as the text appears in both plots.}
    \label{fig:mean_rms_gr_cy}
\end{figure}
Figure \ref{fig:mean_rms_gr_cy} summarizes the values of the mean and RMS velocities at all FOVs across all experimental conditions. Apart from providing further evidence of low mean flows and flow homogeneity ($\overline{u_\rms}$ and $\overline{w_\rms}$ are near identical at all FOVs and their spatial variations is small compared their spatial means), the data also show that RMS velocities are functions of the parameters of the sunbathing algorithm ($\mu_\on$ and $\phi$) and the attachments at the jet exit. Larger $\phi$ and $\mu_\on$ generate turbulent flow with larger RMS velocities, as previously found \citep{variano2008random, carter2016generating}, and the cylindrical attachment weakens the RMS velocities. 

Simple scaling arguments would suggest that the RMS velocities would scale with the jet velocity $(u_\rms, \, w_\rms) \sim U_J$ \citep{dou2016piv, hoffman2021isotropic, tan2023scalings}. This expectation is indeed borne out in the fact that the cylindrical attachment reduces the RMS velocities. However, the simple scaling does not explain the fact that the RMS velocities are sensitive to $\phi$ and $\mu_\on$ for random-jet-stirred flow. \citet{variano2008random} present arguments for the non-monotonic dependency on $\phi$ for a single jet array to the effect that the optimal source fraction is the one where the mean shear is maximised in the `temporary states' created by the sunbathing algorithm. It wasn't clear to us \textit{a priori} whether the optimal value of $\phi$ might perhaps be lower with four jet arrays instead of one, but the data in figure \ref{fig:mean_rms_gr_cy} show that $\phi = 12.5 \%$ produces higher RMS velocities than $\phi = 6.25 \%$, suggesting that the previously found value of $\phi_{\textrm{optimal}} = 12.5 \%$ is robust to additional jet arrays. 

Our data (and previous works discussed in Sec.~\ref{sec:background}) show that RMS velocities increase with $\mu_\on$ up to a saturation point. To better understand this better, we return to the results in Sec.~\ref{sec:singlejetdata}, where it was found that for jets operated in pulsed mode the velocity in the tank centre increased with increasing on-time up to a value of $T_\on U_J/ (\tfrac{1}{2}L_A) = 5$ when it reached the saturation value of a continuous jet. This clearly suggests that $\mu_\on U_J/ (\tfrac{1}{2}L_A)$ is the relevant dimensionless mean on-time and that the RMS velocities are likely to saturate beyond $\mu_\on U_J/ (\tfrac{1}{2}L_A) \gtrsim 5$. Figure \ref{fig:mean_rms_gr_cy} provides reasonable support for this argument, particularly for ${w_\rms}$ which is a better measure of the intensity of turbulent motions since vertical motions are less contaminated by the jet flow. The vertical RMS velocities appear to saturate for the grid attachment for $\mu_\on \ge 3$ s which is equivalent to $\mu_\on U_J/ (\tfrac{1}{2}L_A) \gtrsim 5$. 
%
\subsubsection{Scales of motion: integral scale and Taylor scale}\label{sec:scalesofmotion_length}
To estimate the different lengthscales of turbulence, we use the two-point autocorrelation
\begin{equation}
    \rho_{ij} (r) = \frac{\langle u^\prime_{i}(\bm{x}) \cdot u^\prime_{i}(\bm{x}+r\bm{e_{j}})\rangle}{u_{i,\rms}^2},
    \label{eq:autocorr}
\end{equation}
where $r$ is the spatial lag in the direction given by the unit vector $\bm{e_j}$. This computation is conducted in the homogeneous turbulence region. 

Before analyzing the turbulence scales, we check that combining data from the two cameras does not introduce significant errors. The discrepancy between the autocorrelation functions in the homogeneous region measured by Camera 1 ($-10 < x < 0$ cm) and by Camera 2 ($0 < x <10$ cm) are less than 5\% different from the autocorrelation function calculated by combined data from Camera 1 and 2 ($-10 < x < 10$ cm). Thus, we use the velocity fields from both cameras to compute the autocorrelation, which expands the spatial lag in the $x$ direction.%
\begin{figure}
    \centering
    \includegraphics[width=\textwidth]{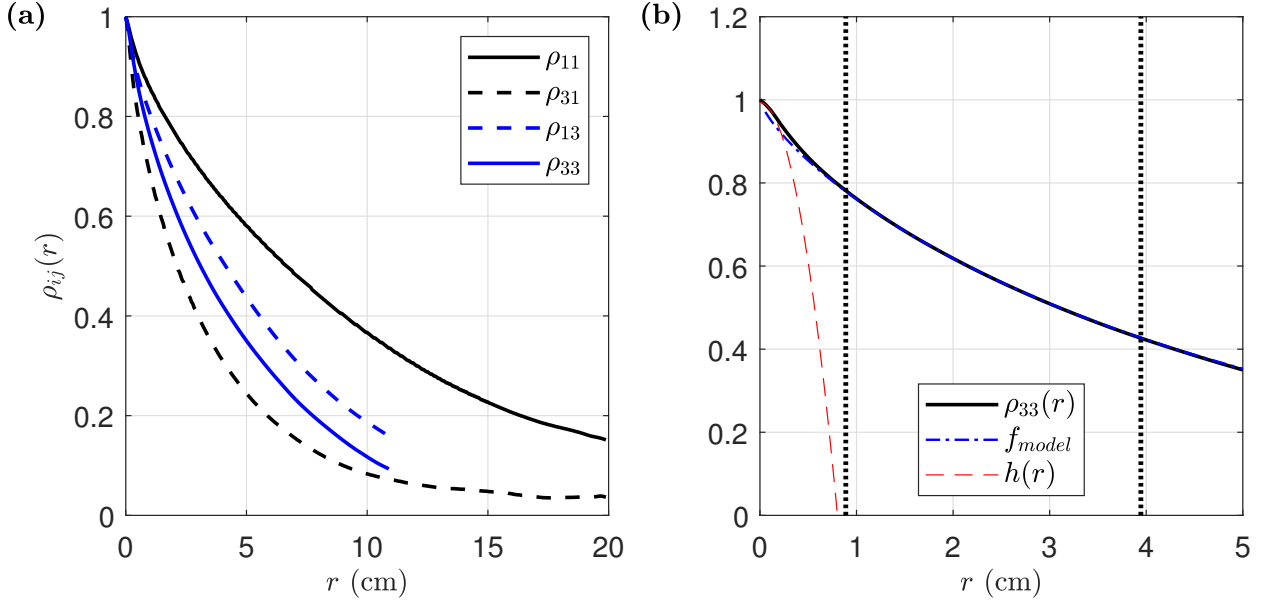}
    \caption{Autocorrelation functions for the GrHi30 case at FOV2 (a) and the $z$ direction longitudinal autocorrelation function with the model function for the inertial subrange and the parabolic fit at the origin (b). The dotted lines are the lower and upper bound of the inertial subrange.}
    \label{fig:autocorr_GrLo30}
\end{figure}

Figure \ref{fig:autocorr_GrLo30}a shows the autocorrelation functions for the GrHi30 case at FOV2. We can see that despite combining data from both cameras, we do not have the required measurement volume to measure the autocorrelation to its first zero-crossing and evaluate the integral length scale directly as $L_{ij}= \int_{0}^{\infty} \rho_{ij}(r) dr $. Instead, we estimate the integral length scales by fitting a model function to $\rho_{ii}$ in the inertial subrange
\begin{equation}
    f_{\textrm{model}} \left(\frac{r}{L^M}\right) = \frac{2}{\Gamma(q)} \left( \frac{r}{2L^M} \alpha \right)^{q} K_{q} \left(\frac{r}{L^M} \alpha \right)
    \label{eq:Model_func}
\end{equation}
where $\alpha = \sqrt{\pi}\Gamma(q+\frac{1}{2})/\Gamma(q)$, $\Gamma$ is the gamma function, and $K_q$ is the modified Bessel function of the second kind \citep{pope2000turbulent}. The model function has two fitting parameters, $q$ and $L^M$, where the subscript $M$ refers to the fact the integral scale is computed fitting data to the model Eq.~(\ref{eq:Model_func}). Figure \ref{fig:autocorr_GrLo30}b shows that the model function fits $\rho_{11}$ very well in the inertial subrange (determined from plateaus of the compensated second-order structure functions as described below). 
\begin{table}[b!]
\caption{Two-point turbulence statistics in the homogeneous region at FOV2}
\begin{adjustbox}{width=1\textwidth}
\small
\centering
\begin{tabular}{|c|c c c c c c|c|c|c|c c c|c c|} 
\hline
 & $L^M_{11}$ & $L^M_{33}$ & $L^F_{11}$ & $L^F_{31}$ & $L^F_{13}$ & $L^F_{33}$ & $\lambda_{z}$ & $T_{z}$ & $\langle \epsilon \rangle$ & $\eta$& $u_{\eta}$& $\tau_{\eta}$ & $Re_{\lambda}$ & $Re_L$\\
\hline
 & & & & & & & & & $(\times 10^{-3})$ &  &  &  & $(\times 10^{2})$ & $(\times 10^{3})$ \\
Case & \multicolumn{6}{c|}{cm} & mm & s & m$^2$/s$^3$ & mm & mm/s & ms &  &  \\
\hline
GrLo15 & 8.35 & 4.15 & 7.83 & 3.13 & 4.78 & 3.85 & 6.73 & 0.86 & 1.19 & 0.155 & 5.69 & 27.3 & 3.42 & 1.96 \\
GrLo30 & 11.03 & 4.96 & 9.79 & 3.51 & 5.90 & 4.62 & 6.79 & 0.93 & 1.43 & 0.148 & 5.96 & 24.9 & 3.82 & 2.61 \\
GrLo60 & 14.27 & 5.31 & 11.89 & 3.79 & 6.54 & 5.02 & 7.01 & 1.00& 1.33 & 0.150 & 5.89 & 25.5 & 3.97 & 2.85 \\
GrHi15 & 7.33 & 4.15 & 7.27 & 2.94 & 4.41 & 3.94 & 6.48 & 0.83 &1.42 & 0.149 & 5.95 & 25.0 & 3.47 & 2.11\\
GrHi30 & 9.17 & 5.01 & 8.30 & 3.69 & 5.41 & 4.83 & 6.35 & 0.82 &2.28 & 0.132 & 6.71 & 19.7 & 4.23 & 3.22\\
GrHi60 & 11.43 & 5.86 & 10.01 & 4.69 & 6.78 & 5.47 & 6.56 & 0.86&2.52 & 0.129 & 6.87 & 18.7 & 4.73 &3.95 \\
\hline 
CyLo15 & 8.1 & 4.58 & 7.26 & 3.26 & 4.50 & 4.41 &8.44 & 1.5 & 0.32 & 0.215 & 4.11 & 52.4 & 2.80 & 1.47 \\
CyLo30 & 11.44 & 5.75 & 9.82 & 4.42 & 6.01 & 5.34 & 8.27 & 1.47&0.51 & 0.191 & 4.62 & 41.5 & 3.40 & 2.20\\
CyLo60 & 13.76 & 6.42 & 11.72 & 4.40 & 7.42 & 5.72 & 8.44 & 1.49&0.55 & 0.188 & 4.70 & 40.1 & 3.66 & 2.48 \\
CyHi15 & 6.92 & 5.22 & 6.72 & 3.29 & 4.36 & 4.44 & 8.29 & 1.47&0.35 & 0.211 & 4.20 & 50.1 & 2.83 & 1.52\\
CyHi30 & 9.45 & 5.70 & 8.50 & 4.03 & 5.64 & 5.02 & 7.97 & 1.30&0.63 & 0.182 & 4.85 & 37.6 & 3.48 & 2.20 \\
CyHi60 & 11.35 & 7.18 & 10.27 & 4.68 & 6.46 & 5.63 & 7.83 & 1.29&0.82 & 0.170 & 5.20 & 32.8 & 3.86 & 2.77\\
\hline
\end{tabular}
\end{adjustbox}
\label{table:2pt_TurbulenceStatisticsFOV2}
\end{table}

For completeness, we also calculate the integral length scales by fitting an exponential curve to the autocorrelation functions, $f_{\textrm{exp}} = \exp{(-x/L^F_{ij})} $. The superscript F refers to the fact that the integral scale is calculated from a fit the exponential function. The differences between $L^M_{ii}$ and $L^F_{ii}$ are less than 15\%.
\begin{figure}
    \centering
    \includegraphics[width=0.75\textwidth]{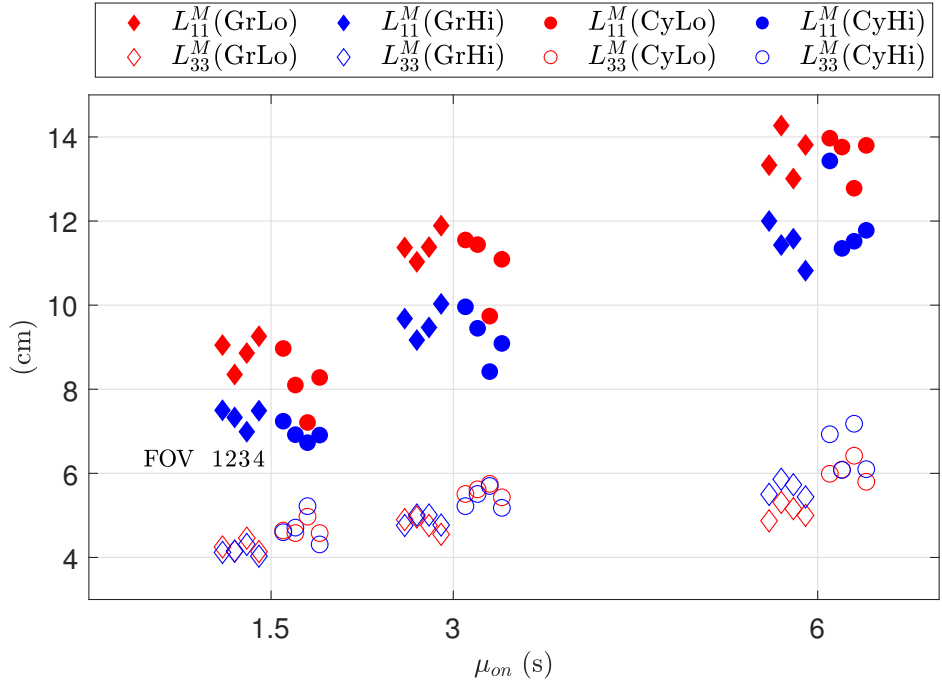}
    \caption{Longitudinal integral length scales estimated by the model function as a function of $\mu_\on$. Quantities for the grid and cylinder cases are located left and right sides of the corresponding $\mu_\on$, respectively. Those for different FOVs are presented in an offset manner (left to right) at the corresponding position.}
    \label{fig:Lint_gr_cy}
\end{figure}

Figure \ref{fig:Lint_gr_cy} and table \ref{table:2pt_TurbulenceStatisticsFOV2} summarize the various integral scales. We observe again that the values are very similar at different FOVs for the same experimental condition which shows the turbulence is homogeneous. Across all FOVs and experimental conditions, $L^M_{11}$ is always larger than $L^M_{33}$ and $L^F_{11}>L^F_{13}>L^F_{33}>L^F_{31}$, which is consistent with the large-scale anisotropy discussed above. Additionally, lower $\phi$ and/or higher $\mu_\on$ yield larger $L^M_{11}$, while $L^M_{33}$ appears to be insensitive to $\phi$ and only a function of $\mu_\on$. The cylindrical attachment has little effect on $L^M_{11}$, but does increase $L^M_{33}$ slightly. 

The simplest scaling argument to predict the integral scale states that $L \sim D_J$ \citep{carter2016generating, hoffman2021isotropic, tan2023scalings} since all scales of motion in a canonical jet scale with the jet diameter. Focusing on $L_{33}$ as the most representative measure of the largest turbulent motions, we see evidence to support this argument since the cylindrical attachment, which widens the jet and creates a larger effective jet diameter, consistently produces a larger integral scale compared to the grid attachment. The scaling of the integral scale with the (effective) jet diameter is also supported by the fact that $L$ is insensitive to changes in $\phi$. However, this scaling argument does not predict that the integral scale also depends on $\mu_\on$. We also expect $L_A$ to be important \citep[][]{carter2016generating, tan2023scalings} and that the dependencies on $\mu_\on$ and $L_A$ to be linked via $\mu_\on U_J/ L_A$. The data in figure \ref{fig:Lint_gr_cy} support $L \sim D_J (\mu_\on U_J/ L_A)$ where the dependency on mean on-time again saturates for $\mu_\on U_J/ (\tfrac{1}{2}L_A) \gtrsim 5$.

\begin{figure}
    \centering
    \includegraphics[width=0.7\textwidth]{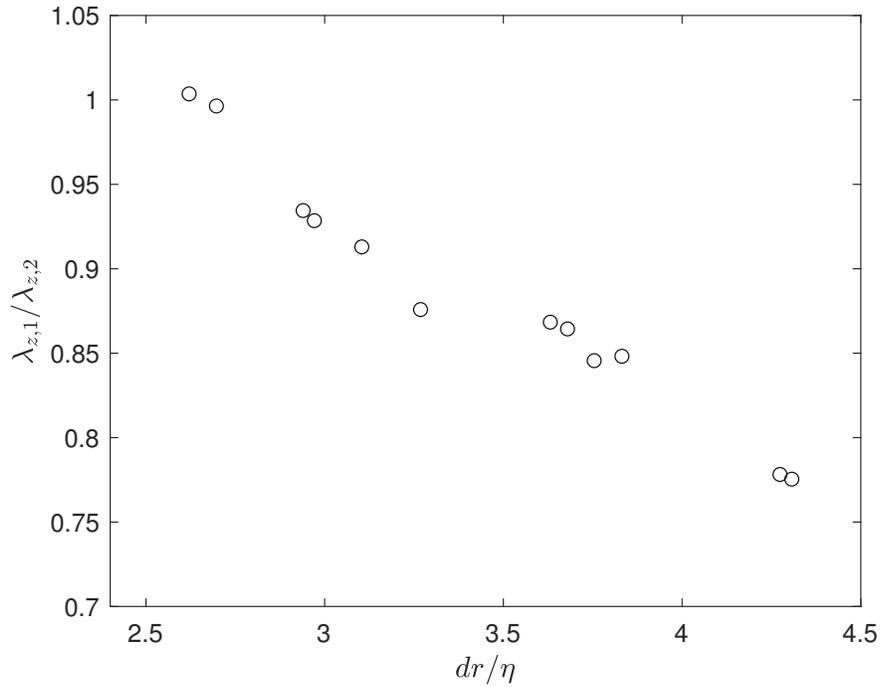}
    \caption{Ratio of $\lambda_{z}$ calculated using Eq.~(\ref{eq:Taylor_2}) ($\lambda_{z1}$) and by fitting the Eq.~(\ref{eq:parabola_fit}) to the longitudinal autocorrelation function ($\lambda_{z2}$) as a function of $dr/\eta$. The data are averaged across all FOVs.} 
    \label{fig:lambda_vs_dr_eta}
\end{figure}
The Taylor microscale $\lambda$ can be calculated from its definition by by fitting an osculating parabola 
\begin{equation}
    h(r) = 1-r^2/\lambda^2,   
    \label{eq:parabola_fit}
\end{equation}
to the autocorrelation function near the origin or it can be calculated via the relation
\begin{equation}
    \lambda_{z} = w_\rms \sqrt{30\nu/\epsilon}, 
    \label{eq:Taylor_2}
\end{equation}
by assuming small-scale isotropy \citep{pope2000turbulent}. We follow both methods, where we fit Eq.~(\ref{eq:parabola_fit}) to the first 2 points of the autocorrelation function excluding the point at $r=0$ (figure \ref{fig:autocorr_GrLo30}b), and compare the results from both methods as a function of the spatial resolution of the data in figure \ref{fig:lambda_vs_dr_eta}. While both methods yield the same answer when the flow field is resolved to approximately $2.5\eta$ (where $\eta$ is the Kolmogorov microscale and discussed more below), fitting a parabola to the autocorrelation data overestimates the Taylor scale when the data is less well resolved. Hence, we report $\lambda_{z}$ calculated from Eq.~\ref{eq:Taylor_2} using FOV2 data in table \ref{table:2pt_TurbulenceStatisticsFOV2}.

\subsubsection{Scales of motion: inertial subrange and dissipation scales}\label{sec:scalesofmotion_equilibriumrange}
To examine the flow statistics in the inertial and dissipation subranges, we use the second-order structure functions 
\begin{equation}
    D^2_{ij}(r) = \langle (u^\prime_{i}(\bm{x}) - u^\prime_{i}(\bm{x}+r\bm{e_{j}}))^2\rangle,
    \label{eq:2nd_str_ftn}
\end{equation}
computed with data from both cameras in the homogeneous turbulence region. 
\begin{figure}
    \centering
    \includegraphics[width=1\textwidth]{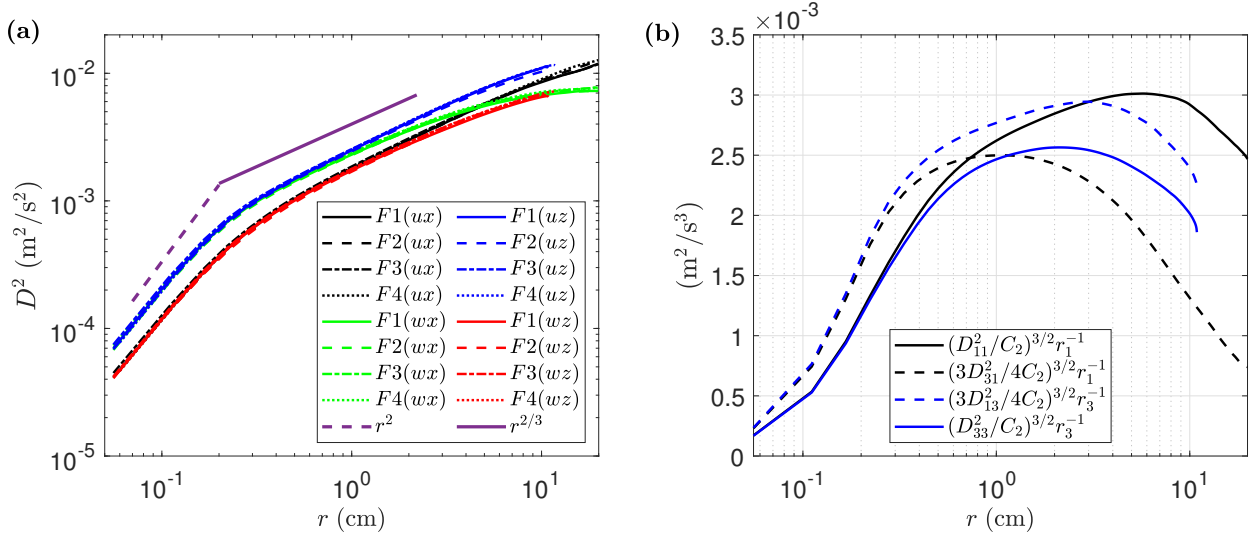}
    \caption{Second-order structure functions for the GrHi60 case across different FOVs (a) and the corresponding compensated structure functions at FOV2 (b). The components of velocity and direction for calculating the second-order structure function using Eq.~(\ref{eq:2nd_str_ftn}) are noted in the parenthesis in the legend of (a).}
    \label{fig:Str_GrHi60}
\end{figure}

Figure \ref{fig:Str_GrHi60} shows the second-order structure functions in the GrHi60 case, which serves as a representative example for all cases for scaling behaviour. We see that the data show the expected power-law scaling in the dissipation range ($\sim r^2$) and in the inertial subrange ($\sim r^{2/3}$). We can also extract information on the degree to which the flow is homogeneous and isotropic across scales. For this, the GrHi60 case is a worst-case example since it has the highest $\mu_\on$. In terms of homogeneity, figure \ref{fig:Str_GrHi60}a shows that there is little-to-no variation between data at different FOVs (lines of the same colour overlap with each other), allowing us to conclude that the flow is homogeneous across all scales. In terms of isotropy, figure \ref{fig:Str_GrHi60}a shows that the longitudinal ($D^2_{11}$ and $D^2_{33}$) and transverse ($D^2_{31}$ and $D^2_{13}$) structure functions calculated using different velocity components overlap with each other, respectively, for $r \lesssim 1$ cm and do no deviate too far from each other for $r \lesssim 3$ cm. In figure \ref{fig:Str_GrHi60}a, which shows the compensated structure function data at FOV2, we can see more clearly that the dissipation range scales are very isotropic and that the inertial subrange scales are reasonably isotropic. Clearly, the large scale flow which has anisotropic features is becoming more isotropic down the turbulence energy cascade, as expected from classical Kolmogorov theory. 

To identify the extent of the inertial subrange and also estimate the energy dissipation rate, we apply the Kolmogorov similarity hypotheses which predict that the longitudinal and transverse structure functions are uniquely determined by $\langle \epsilon \rangle$
\begin{subequations} \label{eq:2ns_str_ftn_comp}
\begin{align}
    D^2_{ii}(r)=C_2(\langle \epsilon \rangle r_i)^{2/3}, \\
    D^2_{ij}(r)=\frac{4}{3}C_2(\langle \epsilon \rangle r_j)^{2/3},
\end{align}    
\end{subequations}
with $C_2 = 2.0$ being the Kolmogorov constant \citep{pope2000turbulent}. From the compensated structure functions (such as the GrHi60 case at FOV2 shown in figure \ref{fig:Str_GrHi60}b), we identify the extent of the inertial subrange as the range where the compensated structure function is within 5\% of its maximum. Figure \ref{fig:Str_GrHi60}b shows that we obtain a much cleaner and clearer plateau in the $D^2_{33}$ data compared with the $D^2_{11}$ data, which show a narrow plateau at larger scales that overlap with our estimates of the integral scale. We interpret this as suggesting that the flow forcing from the jets pumps extra energy into the $u$ component of velocity at scales within the turbulent energy cascade.

To evaluate $\langle \epsilon \rangle$, we use the compensated $D^2_{33}$ data and invert the relation in Eq.~(\ref{eq:2ns_str_ftn_comp}). These values of $\langle \epsilon \rangle$ are summarized in table \ref{table:2pt_TurbulenceStatisticsFOV2}. As a check of the sensitivity, we also compute $\langle \epsilon \rangle$ using $D^2_{31}$ data and the standard formulations
\begin{equation} \label{eq:scaling_dssp}
    \langle \epsilon \rangle = C \frac{u_\rms^3}{L^F_{11}}, \quad\quad \langle \epsilon \rangle = C \frac{w_\rms^3}{L^F_{33}}
\end{equation}
where C = 0.5.

\begin{figure}
    \centering
    \includegraphics[width=0.7\textwidth]{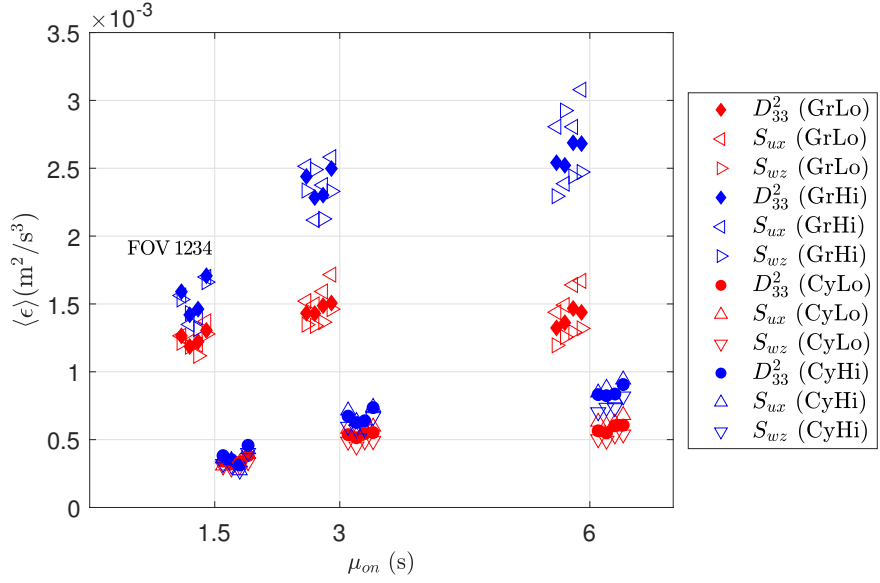}
    \caption{Summary of mean dissipation rate estimated from the structure function ($D^2_{33}$) and the scaling method, denoted as $S$, as a function of $\mu_\on$. Quantities for the grid and cylinder cases are located left and right sides of the corresponding $\mu_\on$, respectively. Those for different FOVs are presented in an offset manner (left to right) at the corresponding position. Subscripts of $S$ represent the velocity and coordinate components chosen for the calculation. The grid and cylinder cases are offset from corresponding $\mu_\on$.}
    \label{fig:dssp_gr_cy}
\end{figure}

Figure \ref{fig:dssp_gr_cy} presents a comparison of $\langle \epsilon \rangle$ under different experimental conditions. There is clear evidence for homogeneity in how similar the values of $\langle \epsilon \rangle$ are across different FOVs. Additionally, the values obtained from the $D^2_{33}$ data closely match the values obtained from Eq.~(\ref{eq:scaling_dssp}) (figure \ref{fig:dssp_gr_cy}) and the values obtained from the $D^2_{31}$ (not shown).

The data in figure \ref{fig:dssp_gr_cy} also show that $\langle \epsilon \rangle$ increases with higher $\phi$ and/or longer $\mu_\on$, with the cylinder attachment decreasing the dissipation rate relative to the grid attachment. Since the data show that the scaling relationship in Eq.~\ref{eq:scaling_dssp} accurately predicts the dissipation rate, we can extrapolate understanding of how RMS velocities and integral scales vary with $\phi$, $\mu$, and the attachments to understand how the dissipation rate varies with these parameters. The main finding is again that the dissipation rate can be increased by increasing the mean on-time, but the effect saturates for $\mu_\on U_J/ (\tfrac{1}{2}L_A) \gtrsim 5$.

Using the estimated $\langle \epsilon \rangle$ from $D^2_{33}$ data, the Kolmogorov microscales are calculated via the relations 
\begin{equation} \label{eq:Kolmogorov_scales}
    \eta = (\nu^3/\langle \epsilon \rangle)^{1/4}, \quad\; u_{\eta} = (\nu \langle \epsilon \rangle)^{1/4}, \quad\; \tau_{\eta} = (\nu/\langle \epsilon \rangle)^{1/2}, 
\end{equation}
where $\eta, u_{\eta}$ and $\tau_{\eta}$ are the Kolmogorov length, velocity, and time scales, respectively. These values are reported in table \ref{table:2pt_TurbulenceStatisticsFOV2}. 
%
\subsubsection{Scale separation as a function of Reynolds number} \label{sec:scaleseparation}
\begin{figure}
    \centering
    \includegraphics[width=0.6\textwidth]{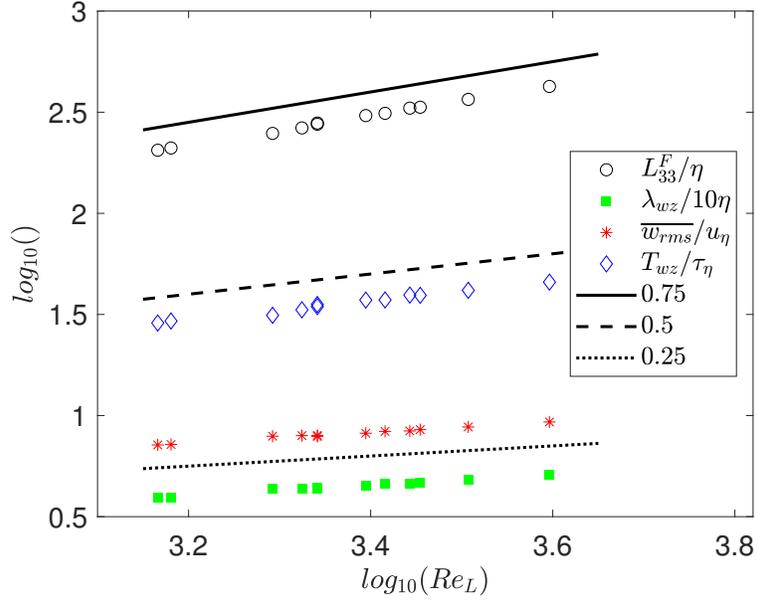}
    \caption{Largest to smallest scale ratio regarding Reynolds number at FOV2. Values are shown as log scale to illustrate the exponential scaling relation. Note, the values for $\lambda_{z}/\eta$ are plotted a decade smaller for easier visualization.}
    \label{fig:scale_ratio_vs_Re_L}
\end{figure}
High Reynolds number turbulent flows have a large separation between the largest and smallest scales. We inspect this dependency in our data by comparing it with the scaling laws
\begin{equation} \label{eq:ReL_scalings}
        L_{0}/\eta \sim Re_L^{3/4}, \quad\; \lambda/\eta \sim Re_L^{1/4}, \quad\; u_{0}/u_{\eta} \sim Re_L^{1/4}, \quad\; T_{0}/\tau_{\eta} \sim Re_L^{1/2},  
\end{equation}
where $L_{0}, u_{0}$, $T_{0}$ are the largest characteristic scales of length, velocity, and time, respectively, and $Re_L = u_0 L_0 / \nu$ is the Reynolds number based on these scales \citep{pope2000turbulent}. Using $L_{0} = L^F_{33}$, $u_{0} = \overline{w_\rms}$, and $T_{0} = L^F_{33}/\overline{w_\rms}$, this Reynolds number becomes $Re_L = \overline{w_\rms} L^F_{33}/\nu$. Figure \ref{fig:scale_ratio_vs_Re_L} shows the comparison of the ratio of the largest to smallest scales as a function of $Re_L$ for all experimental conditions at FOV2, where we observe that the data show a very high degree of correlation with the expected scaling laws ($R^2$ values greater than 0.95).

\begin{figure}
    \centering
    \includegraphics[width=1\textwidth]{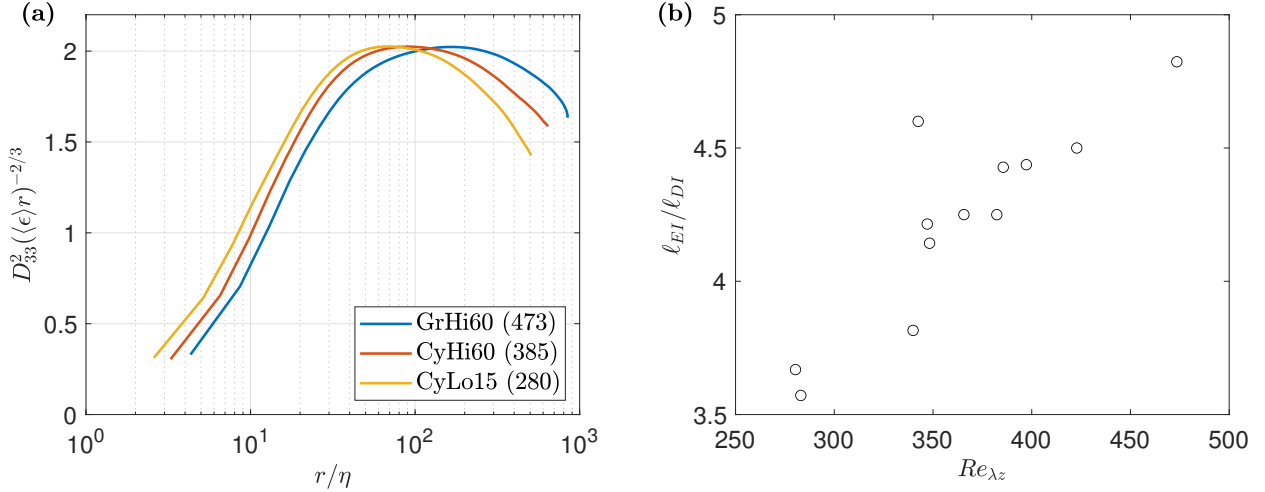}
    \caption{Normalised $D^2_{33}$ for different $Re_{\lambda z}$ (a), and the extent of the inertial subrange as measured by $\ell_{EI}/\ell_{DI}$ as a function of $Re_\lambda$ (b). $\ell_{EI}$ and $\ell_{DI}$ are the demarcation lines between energy-containing and inertial ranges and between inertial and dissipative ranges, respectively.}
    \label{fig:D33_vs_Re_lambda}
\end{figure}
In idealised turbulence studies, the Taylor scale Reynolds number is more commonly used. We define it here as $Re_{\lambda} = \overline{w_\rms} \lambda_{z}/\nu$, choosing the vertical characteristic velocity and length scales as before. $Re_{\lambda}$ falls in the range $340$--$470$ for the grid attachment, and $280$--$380$ for the cylinder attachment (table \ref{table:2pt_TurbulenceStatisticsFOV2}). While it is common for laboratory studies to report a Taylor scale Reynolds number, it is often difficult to directly compare Reynolds number effects across studies since different stirring mechanisms and different methods for calculating the Reynolds number can lead to differences in the scale separation for the same Reynolds number. Thus, in figure \ref{fig:D33_vs_Re_lambda}a we show the compensated longitudinal structure function for different cases and compute the scale separation observed in the inertial subrange as before. Figure \ref{fig:D33_vs_Re_lambda}b shows how this scale separation varies with our reported $Re_{\lambda}$. We find that our scale separation is very similar in magnitude to standard DNS of turbulence in a cubic periodic box \citep{Ishihara2009}, which makes it easier to directly compare studies in similar Reynolds numbers across different lab studies and DNS.
%
\section{Discussion} \label{sec:discussion}
Within the context of a new facility, where we generate random-jet-stirred turbulence in a vertical octagonal prism-shaped tank using four jet arrays on four faces of the tank, we have studied high-Reynolds-number turbulence with a negligible mean flow and mean shear that is homogeneous over a large domain (2.5--4$L$ in the horizontal direction and 5--8$L$ in the vertical direction). The tank design ensures that the flow is isotropic in horizontal planes due to jet forcing from four directions, but this dominance of forcing in the horizontal direction also leads to anisotropy at largest scales and an excess of horizontal momentum in the turbulent scales of motion. By investigating the properties of the turbulence at different scales, we find that anisotropy at large scales decays to produce scale-local isotropic turbulence within the inertial subrange and dissipation range, with the isotropy improving as energy moves down scale. We can also control the Reynolds number and scales of turbulent motion (RMS velocities, integral scales, and dissipation rate) to varying degrees by changing the parameters of the jet driving algorithm. By examining the ratio of integral-to-Kolmogorov scales and span of the inertial subrange, we confirm that the flow statistics obey the expected scaling relationships for scale separation as a function of Reynolds number for idealised turbulence.

As noted in Sec.~\ref{sec:introduction}, laboratory facilities using randomized stirring from multiple units (e.g., jets, impellers) produce turbulence that has a smaller mean flow and better homogeneity and isotropy over a larger region compared with facilities that use continuous stirring, usually at the cost of lower turbulence intensity and Reynolds number. Additionally, the complexity introduced by randomized stirring also means that it becomes more difficult to predict the scales of turbulence as a function of the flow from each stirring unit, tank size, and the algorithm that drives the units. In this regard, we have made a number of steps forward within the context of random-jet-driven turbulence using the sunbathing algorithm. 

While most facilities use one or two facing arrays, we have found that the addition of two more arrays has several benefits. We are able to achieve a higher turbulence intensity and Reynolds number with improved isotropy. We are also able to reduce the ``jet merging'' distance (defined here as the minimum distance from each array where the turbulence is homogeneous) from $L_\JM \approx 6 J$ (where $J$ is the inter-jet spacing) as previously found to $L_\JM \approx 4 J$. This is due to the arrangement of jet arrays that are perpendicular to each other and therefore merge with the surrounding flow earlier. This can be important since the jet merging distance is one of the considerations when designing the tank size. In terms of the optimal source fraction $\phi$ (the fraction of jets that are on at any given time), we found that the addition of more jet arrays does not change the optimum value; it is within the range 5--25\% as originally found with a single jet array.

We also tested manipulating the properties of each jet by introducing 3D-printed attachments to the jet exit, including a grid attachment to provide flow conditioning and a horizontal cylindrical attachment to induce unsteady flow separation and additional vertical mixing. By examining the flow from an individual jet in the near-field and far-field, we observed that the grid attachment successfully conditioned the flow so that the jet flow was close to a canonical non-swirling turbulent round jet. The cylindrical attachment induced a wake in the near-field that likely introduced more vertical mixing, but in the far field the flow relaxed to produce a wider and slower jet profile. This resulted in lower RMS velocities for the cylindrical attachment. Notably, the ratio of the normal fluctuations ($w_\rms$) to the axial fluctuations ($u_\rms$) was similar for both attachments in the far field, and thus, the cylindrical attachment yielded only modest gains in flow isotropy in the tank centre. Interestingly, we also found that the integral scale increased slightly for the cylindrical attachment, which we attribute to a wider effective jet diameter. Overall, we find that attachments at the jet exit can help to reduce the effective jet exit velocity and increase the effective jet diameter, but due to turbulent mixing, near-field changes in the flow are not necessarily retained in the far field.  

In random-jet-stirred turbulence, the sunbathing algorithm drives the jets in pulsed mode with a mean on-time $\mu_\on$, but this mean on-time has usually been reported in dimensional terms and the nature of the pulsed-jet flow had not been previously investigated. We introduced two possible dimensionless mean on-times: (1) based on the jet exit velocity and jet diameter, $\mu_\on U_J/D_J$; and (2) based on the jet exit velocity and tank size, $\mu_\on U_J / (\tfrac{1}{2}L_A)$ where $L_A$ is the distance between facing jet arrays. Measurements of the near-field flow from an individual jet operated in pulsed mode with an on-time corresponding to $\mu_\on$ showed that there is always a starting vortex, but that for $\mu_\on U_J/D_J \gtrsim 4$, there is a trailing jet as previously found \citep{gharib1998universal}. Thus, we recommend that $\mu_\on U_J/D_J \approx 4$ be used as the lower bound for selecting the mean on-time. Correspondingly, the far-field measurements of an individual pulsed jet showed that the momentum of the fluid set into motion diffuses and thus the peak velocity measured at the tank centre is smaller compared to a continuous jet. However, as the on-time increases, the velocity of a pulsed jet recovers to the continuous jet value for $T_\on U_J / (\tfrac{1}{2}L_A) \gtrsim 5$. Increasing the on-time beyond this value does not provide a further increase in jet velocity at the tank centre. Thus, we recommend that $\mu_\on U_J/(\tfrac{1}{2}L_A) \approx 5$ be used the upper bound for selecting the mean on-time. The saturation of turbulence scales with increasing mean on-time also has a weak dependency on the source fraction $\phi$, which is more complex and requires further investigation. However, it is clear an upper bound on $\mu_\on U_J / (\tfrac{1}{2}L_A)$ explains the saturation of turbulence scales with increasing mean on-time observed in previous studies \citep{variano2008random, perez2016effect, carter2016generating} and corroborates the intuitive argument that increasing mean on-time beyond a certain threshold only increases the strength of the secondary circulations and does not increase turbulent kinetic energy. We also confirm this intuition from spectra of velocity fluctuations using data collected at 0.5 Hz and 20 Hz. The data show that the timescales associated with the sunbathing algorithm produce a clear peak in the data when $\mu_\on U_J/(\tfrac{1}{2}L_A) \gtrsim 10$. The timescale at which we observe this peak is $\mu_\on + \mu_\off$ \citep[as also found in][]{lawson2022unsteady}, rather than $\phi \mu_\on$ \citep[as hypothesized in][]{variano2008random}.

\section*{Acknowledgements}
We gratefully acknowledge support from the US National Science Foundation (CBET-2211704). We thank Luke Summey for help with tank design and construction. We also thank Evan Variano and Gautier Verhille for useful discussions.

\bibliographystyle{abbrvnat}
\bibliography{homogeneousturbulencejetarrays}

\end{document}